\documentclass[aps,pra,reprint,amsmath]{revtex4-1}
\usepackage{graphicx} 
\usepackage{hyperref} 
\usepackage{amsfonts} 
\usepackage{xcolor} 

\newcommand\markred[1]{#1}

\renewcommand{\Re}{\mathrm{Re}}
\renewcommand{\Im}{\mathrm{Im}}

\newcommand{\MHz}{\,\mathrm{MHz}}
\newcommand{\GHz}{\,\mathrm{GHz}}
\newcommand{\ns}{\,\mathrm{ns}}
\newcommand{\us}{\,\mathrm{\mu s}}
\newcommand{\dB}{\,\mathrm{dB}}
\newcommand{\W}{\,\mathrm{W}}
\newcommand{\panel}[1]{(\textsf{#1})}

\newcommand{\It}{\tilde{I}}
\newcommand{\Qt}{\tilde{Q}}

\newcommand{\IF}{\mathrm{IF}}
\newcommand{\fs}{f_\mathrm{s}}

\newcommand{\V}{\,\mathrm{V}}

\newcommand{\K}{\,\mathrm{K}}
\newcommand{\mK}{\,\mathrm{mK}}
\newcommand{\m}{\,\mathrm{m}}

\newcommand{\mV}{\,\mathrm{mV}}

\newcommand{\tr}{\mathrm{tr}}
\newcommand{\fb}{\mathrm{fb}}
\newcommand{\fbTime}{\mathrm{fbTime}}

\newcommand{\tauFB}{\tau_\mathrm{FB}}
\newcommand{\tauELtot}{\tau_\mathrm{EL,tot}}
\newcommand{\tauRO}{\tau_\mathrm{RO}}
\newcommand{\tauAP}{\tau_\mathrm{AP}}
\newcommand{\tauAWG}{\tau_\mathrm{AWG}}
\newcommand{\tauGtot}{\tau_\mathrm{G,tot}}

\newcommand{\tauADCDIO}{\tau_\mathrm{ADC,DIO}}
\newcommand{\tauProc}{\tau_\mathrm{proc}}

\newcommand{\Prob}{\mathbb{P}}
\newcommand{\g}{\lvert g\rangle}
\newcommand{\e}{\lvert e\rangle}
\newcommand{\gbra}{\langle g\rvert}
\newcommand{\ebra}{\langle e\rvert}
\newcommand{\Ptherm}{P_{\text{therm}}}

\newcommand{\Mone}{\mathrm{M1}}
\newcommand{\Mtwo}{\mathrm{M2}}

\newcommand{\RGG}{\mathcal{R}_{\mathrm{GG}}}
\newcommand{\RGE}{\mathcal{R}_{\mathrm{GE}}}
\newcommand{\REG}{\mathcal{R}_{\mathrm{EG}}}
\newcommand{\REE}{\mathcal{R}_{\mathrm{EE}}}

\newcommand{\BenADDA}{BenADDA-V4\textsuperscript{\texttrademark}}

\renewcommand{\eqref}[1]{Eq.~(\ref{eq:#1})}
\newcommand{\figref}[1]{Fig.~\ref{fig:#1}}
\newcommand{\secref}[1]{Sec.~\ref{sec:#1}}
\newcommand{\appref}[1]{App.~\ref{app:#1}}
\newcommand{\tabref}[1]{Tab.~\ref{tab:#1}}
\newcommand{\citeref}[1]{Ref.~\cite{#1}}

\newcommand{\Virtex}{Virtex--}

\newcommand{\zurich}{\affiliation{Department of Physics, ETH
Z\"urich, CH-8093 Z\"urich, Switzerland}}

\begin{document}

\title{Low-Latency Digital Signal Processing for Feedback and
Feedforward in\\ Quantum Computing and Communication}
\date{\today}

\begin{abstract}
Quantum computing architectures rely on classical electronics
for control and readout. Employing classical
electronics in a feedback loop with the quantum system allows to
stabilize states, correct errors and to realize specific feedforward-based quantum
computing and communication schemes such as deterministic
quantum teleportation.
These feedback and feedforward operations are required to be
fast compared to the coherence time of the quantum
system to minimize the probability of errors. We present a field
programmable gate array (FPGA) based digital signal processing
system capable of real-time quadrature demodulation,
determination of the qubit state and generation of
state-dependent feedback trigger signals. The feedback trigger
is generated with a latency of $110\ns$ with respect to the
timing of the analog input signal. We characterize the
performance of the system for an active qubit initialization
protocol based on dispersive readout of a superconducting qubit
and discuss potential applications in feedback and feedforward
algorithms.
\end{abstract}

\author{Yves~Salath\'e}
\email{ysalathe@phys.ethz.ch}
\zurich

\author{Philipp~Kurpiers}
\zurich

\author{Thomas~Karg}
\zurich

\author{Christian~Lang}
\zurich

\author{Christian~Kraglund~Andersen}
\zurich

\author{Abdulkadir~Akin}
\zurich

\author{Christopher~Eichler}
\zurich

\author{Andreas~Wallraff}
\zurich

\maketitle


\section{Introduction}

Recent quantum physical research is directed towards gaining
experimental control of large-scale, strongly-interacting
quantum systems such as trapped ions~\cite{Blatt2008} and
solid-state devices~\cite{Girvin2009b}. The ultimate goal is to
realize a quantum
computer~\cite{Kielpinski2002,Devoret2013,Kelly2015,Lekitsch2017}
with a large number of quantum bits (qubits) which may
outperform classical computers for certain computational
tasks~\cite{Nielsen2000,Preskill2012,Ronnow2014,Boixo2017,Jordan2017}.
However, quantum systems do not act as stand-alone components
but must be combined with classical electronics to control
inputs such as microwave pulses or external magnetic fields and
to record and analyze the output signals~\cite{Reilly2015}.
Analyzing the output signals in real time can be advantageous to
condition input signals on prior measurement results and
therefore realize a feedback loop with the quantum
system~\cite{Wiseman2010}.

Quantum feedback schemes~\cite{Zhang2017j} make use of the
results of quantum measurements to act back onto the quantum
state of the system within its coherence time.
Experimental realizations of quantum feedback have shown that it
is possible to prepare and stabilize non-classical states of
electromagnetic fields in optical~\cite{Smith2002,Reiner2004}
and microwave~\cite{Sayrin2011} cavities, and to enhance the
precision of phase measurements using an adaptive homodyne
scheme~\cite{Armen2002}.

The first demonstrations of feedback protocols with
superconducting qubits showed active
initialization of qubits into their ground
state~\cite{Riste2012b} and the stabilization of Rabi and Ramsey
oscillations~\cite{Vijay2012,Campagne-Ibarcq2013}. Further
recent feedback experiments with superconducting qubits
demonstrated the deterministic preparation of entangled
two-qubit states~\cite{Riste2013, Liu2016c}, the reversal of
measurement-induced dephasing~\cite{deLange2014}, and the
stabilization of arbitrary single-qubit states by continuously
observing the spontaneous emission from a
qubit~\cite{Campagne-Ibarcq2016}.

Quantum feedforward schemes are closely related to quantum
feedback schemes. In quantum feedforward schemes one part of a
quantum system is measured while the action takes place on
another part of the quantum system. A prominent example for a
feedforward scheme is the quantum teleportation
protocol~\cite{Bennett1993}, which has been realized with active
feedforward in quantum optics
setups~\cite{Bouwmeester1997,Pittman2002,Giacomini2002,Ma2012},
in molecules using nuclear magnetic
resonance~\cite{Nielsen1998}, trapped
ions~\cite{Barrett2004,Riebe2004}, atomic
ensembles~\cite{Krauter2013} and solid-state
qubits~\cite{Steffen2013,Pfaff2014}.

The feedback latency is commonly defined as the time required
for a single feedback round, i.e.\ the time between the
beginning of the measurement of the state and the completion of
the feedback action onto the state. A general requirement to
achieve high success probabilities in quantum feedback schemes
is that the feedback latency is much shorter than the timescale
on which the quantum state decoheres. 

Analog feedback schemes such as those reported in
Refs.~\cite{Vijay2012,Campagne-Ibarcq2016} feature feedback
latencies on the order of $100\ns$, where the latencies are
limited by analog bandwidth and delays in the cables in the
cryogenic setups. However, analog signal processing circuits
have limited flexibility. The flexibility can be improved by
using a digital signal processing (DSP) unit in the feedback
loop, which can be implemented on a central processing unit
(CPU) or on a field programmable gate array
(FPGA)~\cite{MeyerBaese2014}. CPU-based DSP systems offer
versatile and convenient programming at the cost of several
microseconds latency~\cite{Sayrin2011,Riste2012b} due to the
delays introduced by the digital input and output of the signal,
which is too slow to achieve very low error probabilities for
feedback operations on superconducting qubits.

In this paper, we describe an FPGA-based feedback-capable signal
analyzer which allows for real-time digital demodulation of a
dispersive readout
signal~\cite{Blais2004,Wallraff2005,Gambetta2007} and the
generation of a qubit-state-dependent trigger with
 input--to--output latency of $110\ns$.
Our signal analyzer is therefore among the fastest
feedback-capable digital signal analyzers reported so
far~\cite{Schilcher2008,Campagne-Ibarcq2013,Riste2013,Riste2015a,Ryan2017}.
The capabilities of our signal analyzer enabled the feedforward
action in the deterministic quantum teleportation experiment
presented in \citeref{Steffen2013}. In this paper, we illustrate
the use of the feedback signal analyzer in a feedback loop for
qubit initialization~\cite{Riste2012b} and experimentally
characterize its latency and performance.

The paper is organized as follows: in \secref{overview} we
present an overview of a typical feedback loop in which our
instrument is used and analyze the feedback latency. In
\secref{digital} we discuss the implementation of the digital
signal processing on the FPGA and analyze the processing
latencies. Finally, in \secref{experiment} we experimentally
characterize the performance of the feedback loop. In the
appendices, we provide more details about our experimental setup
and our implementation of the digital signal processing on the
FPGA.


\section{Overview of the feedback loop}\label{sec:overview}

In this section, we explain the elements of a typical feedback
loop shown in \figref{loop}\panel{a}. We designed the feedback
loop to issue pulses onto a superconducting qubit inside a
dilution refrigerator conditioned on a measurement of the qubit
state by analog and digital signal processing using cryogenic
and room-temperature electronics.
We first discuss the elements of the detection scheme and the
actuator electronics and then present the latencies of the
feedback loop. We provide a detailed description of our
experimental setup in \appref{setup}.

\subsection{Principle of the detection scheme}
\label{sec:readoutScheme}

\begin{figure}
    \includegraphics{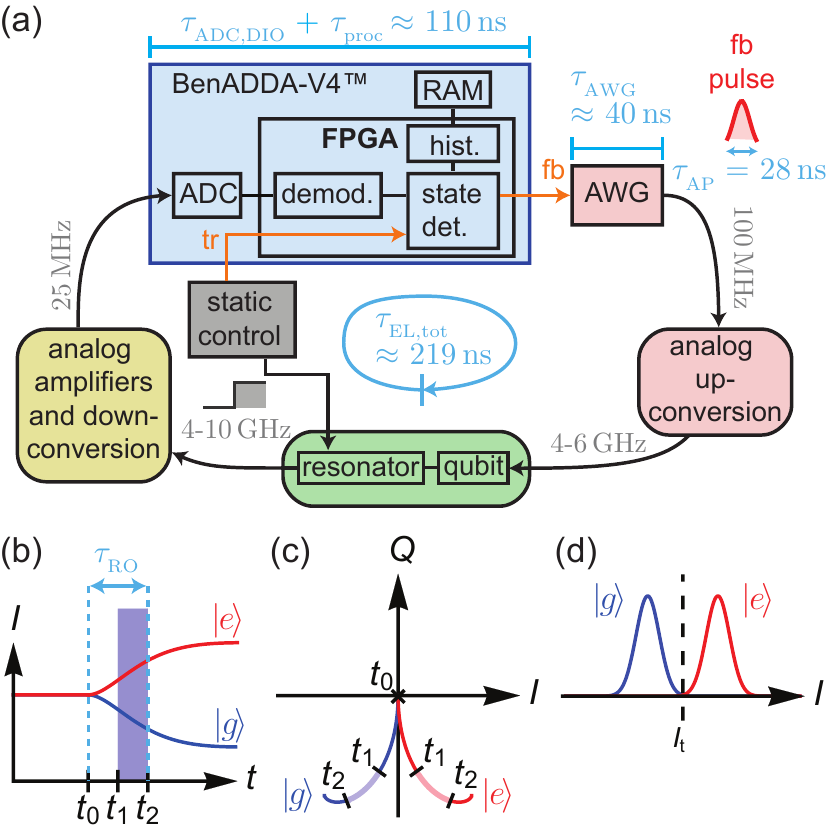}
    \caption{\panel{a} Overview of the feedback loop.
    Typical latencies are indicated in blue and
typical carrier frequencies of the signal are indicated in gray.
See text for details.
\panel{b} Sketch of the time-dependence of the in-phase
component $I$ of the readout signal which approaches different
steady-state values depending on whether the qubit is in state
$\lvert g\rangle$ (blue curve) or $\lvert e\rangle$ (red curve).
We consider a scenario in which the response time of the
resonator is much shorter than the lifetime of the qubit.
Specific times indicated are the onset of the readout pulse
($t_0$) as well as the beginning ($t_1$) and end ($t_2$) of the
integration time ($\tau_\mathrm{i}$, blue shaded region). We
define the total readout time $\tau_\mathrm{RO}$ as the time
difference between $t_0$ and $t_2$ (blue arrow between dashed
lines)~\cite{Walter2017}.
\panel{c} Sketch of the trajectories in the plane spanned by the
$I$ and $Q$ components of the signal for the states $\lvert
g\rangle$ (blue) and $\lvert e\rangle$ (red). Specific points in
the trajectories are marked corresponding to the times $t_0$,
$t_1$ and $t_2$ as defined in \panel{b}.
\panel{d} Sketch of the typical distribution of the integrated
in-phase component ($I$) when the qubit is in state $\lvert
g\rangle$ (blue curve) or $\lvert e\rangle$ (red curve). The
dashed line represents the threshold value $I_\mathrm{t}$ based
on which the state of the qubit is determined.}
    \label{fig:loop}
\end{figure}

We consider the dispersive readout of the state of transmon
qubits~\cite{Koch2007,Schreier2008} with typical frequencies
${\omega_\mathrm{q}/(2\pi)\approx\hbox{4--6}\GHz}$ for the
transition between the ground $\lvert g\rangle$ and first
excited state $\lvert e\rangle$. We couple a microwave resonator
to the qubit [green box in~\figref{loop}\panel{a}] with a
frequency difference between qubit and resonator designed to be
in the dispersive regime~\cite{Blais2004,Wallraff2005}.

In our experimental realization of the feedback
loop~(see~\secref{experiment}), the qubit transition frequency
is ${\omega_\mathrm{q}/(2\pi)=6.148\GHz}$ and the center
resonator frequency amounts to
${\omega_\mathrm{r}/(2\pi)=7.133\GHz}$ with dispersive coupling
rate ${\chi/(2\pi)\approx1.1\MHz}$ between the qubit and the
resonator. Depending on whether the qubit is in state $\lvert
g\rangle$ or $\lvert e\rangle$, we observe the dispersively
shifted resonator frequency at ${\omega_\mathrm{r}\pm\chi}$
respectively.

The qubit-state-dependent frequency shift leads to a state-dependent resonator response when the resonator is
probed with a microwave pulse.
In the dispersive readout scheme, high-fidelity quantum
nondemolition readout~\cite{Wiseman2010} is achieved when
probing the resonator with power ${\kappa \langle
\hat{n}\rangle\hbar\omega_\mathrm{r} \approx 10^{-16}\W}$ such
that the steady-state average photon number $\langle
\hat{n}\rangle$ in the resonator is on the order of 1--10
microwave
photons~\cite{Wallraff2005,Vijay2011,Johnson2011a,Jeffrey2014,Vool2016,Walter2017}.
Due to the low power, it is essential to connect the output of
the resonator to a Josephson parametric
amplifier~(JPA)~\cite{Yurke1989,Yamamoto2008,Castellanos2008,Abdo2011,Vijay2011,Lin2014,Eichler2014,Roy2015,White2015}
to be able to discern the qubit-state-dependent resonator
response within a single repetition of the experiment and in a
time shorter than the qubit lifetime. Other schemes involve the
direct coupling of a qubit to a Josephson bifurcation
amplifier~\cite{Siddiqi2006,Lupascu2007,Mallet2009,Schmitt2014a},
autoresonant oscillator~\cite{Murch2012d} or parametric
oscillator~\cite{Krantz2016} to be able to discern the qubit
state with a higher microwave power.

For simplicity, we consider the case where the resonator is
probed with a microwave pulse with frequency $\omega_\mathrm{r}$
and square envelope. The scheme considered here could be
extended to include more sophisticated pulse
shapes~\cite{Jeffrey2014,McClure2016,Bultink2016,Walter2017}
which increase the speed and fidelity of the readout as well as
the speed of the reset of the intra-resonator field.

We employ the complex representation of the signal ${I(t) +
iQ(t) \equiv A(t)\exp\left[\phi(t)\right]}$ where $A(t)$ and
$\phi(t)$ are the time-dependent amplitude and phase of the
signal at frequency $\omega_\mathrm{r}$. Upon transmission of
the readout pulse with frequency close to resonance, the
time-dependent in-phase $I(t)$ and quadrature $Q(t)$ components
of the signal follow an exponential rise towards steady-state
values starting at time $t_0$ after the onset of the readout
pulse as illustrated in
\figref{loop}\panel{b}~\cite{Bianchetti2009}. The steady-state
values depend on whether the qubit is in state $\lvert g\rangle$
(blue curve) or state $\lvert e\rangle$ (red curve).
The trajectories of the readout signal in the two-dimensional
plane spanned by $I$ and $Q$ as sketched in
\figref{loop}\panel{c} start at the center of the plane which
corresponds to zero amplitude and move into two different
directions depending on the qubit state $\lvert g\rangle$ (blue
curve) or $\lvert e\rangle$ (red curve).

The signal is subject to noise added by
passive and active components~\cite{Gao2011}. Therefore we apply
a linear filter to the signal with the goal to attenuate noise
frequency components while keeping the frequency components that
contain the
signal~\cite{Gambetta2007,deLange2014,Ng2014,Walter2017}. In
particular, we apply a moving average filter which is
advantageous in terms of the signal processing
latency~(see~\secref{DDC}). The moving average is equivalent to
an unweighted integration of the original signal in a particular
integration window starting at a variable time $t_1$ and ending
at time $t_2= t_1+\tau_\mathrm{i}$ [see \figref{loop}\panel{b}
and \figref{loop}\panel{c}], where $\tau_\mathrm{i}$ is a
constant integration time. We define the total readout duration
as the time difference ${\tauRO\equiv t_2-t_0}$ between the
onset of the readout pulse and the end of the integration
window. In the experiment presented in \secref{experiment} we
used an integration window of $\tau_\mathrm{i}=40\ns$ and a
readout duration of \markred{${\tauRO = (105\pm2)\ns}$}.

In the absence of transitions between qubit states during the
integration time, the statistical distribution of the integrated
signal, when the experiment is repeated many times, is expected
to be represented by two Gaussian-shaped peaks in a histogram of
the $I$ component [\figref{loop}\panel{d}]. In the presence of
qubit state transitions during the readout, the distributions
corresponding to the states $\lvert g\rangle$ and $\lvert
e\rangle$ are expected to be non-Gaussian with an increased
overlap~\cite{Gambetta2007,Walter2017}. We discern the states
$\lvert g\rangle$ and $\lvert e\rangle$ of the qubit by
comparing the $I$ signal to a threshold value $I_\mathrm{t}$
[dashed line in \figref{loop}\panel{d}]. The fidelity of the
readout depends on the signal--to--noise ratio of the readout
signal~\cite{Jeffrey2014,Walter2017}. To maximize the readout
fidelity, we optimize the integration window and threshold
value~$I_\mathrm{t}$.

\subsection{Implementation of the detection scheme}
\label{sec:readoutSchemeImpl}

The readout pulse is issued by the static control hardware [gray
box in \figref{loop}\panel{a}]. Simultaneously, the static
control hardware sends a trigger [$\tr$ in
\figref{loop}\panel{a}] to the FPGA to synchronize the digital
signal processing with the readout pulse.

We use an analog detection chain [yellow box in
\figref{loop}\panel{a}] containing amplifiers with a total gain
of approximately $120\dB$~(see \appref{setup}) to detect the
signal at the output of the resonator. In addition, the
detection chain uses analog down--conversion electronics to
convert the readout signal to an intermediate frequency
$\omega_{\mathrm{IF}}$ compatible with the sampling rate
$\fs=100\,\mathrm{MS/s}$ of our DSP unit. We choose an
intermediate frequency at a quarter of the sampling frequency,
i.e.\ $\omega_{\mathrm{IF}}/(2\pi)=\fs/4=25\MHz$, which allows
for efficient digital down--conversion~(see \secref{DDC}). Note
that in principle it is possible to directly demodulate the
signal into its $I$ and $Q$ components in the analog signal
processing but this requires the $I$ and $Q$ component of the
signal to be digitized using two separate analog-to-digital
converter (ADC)
channels~\cite{Considine1983,Schilcher2008,Lyons2011}. The
separate digitization of the $I$ and $Q$ components is sensitive
to mismatches between the conversion-loss and reference level
which lead to a distortion of the digitized complex signal. In
contrast, down--conversion to an intermediate frequency in the
range of $10\MHz$ to $1\GHz$ avoids low-frequency noise, DC
offsets and requires only one ADC channel at the cost of a
reduced bandwidth~\cite{Considine1983,Schilcher2008,Lyons2011}.

We implement the digital signal processing on a Xilinx
\hbox{\Virtex4} FPGA mounted on a commercial DSP unit by
Nallatech, Inc. (\BenADDA) [blue box in \figref{loop}\panel{a}]
which includes an ADC with sampling rate
$\fs=100\,\mathrm{MS/s}$ and 14-bit voltage resolution.
In a first step, the DSP digitally demodulates the signal
[labeled as demod.\ in \figref{loop}\panel{a}].
The
state discrimination module [state det.\ in
\figref{loop}\panel{a}] then compares the filtered $I$ signal at time
$\tauRO$ to the threshold $I_\mathrm{t}$, to determine the qubit
state from the demodulated signal.
Depending on the determined qubit state, a feedback trigger [$\fb$
in \figref{loop}\panel{a}] is sent from the FPGA to the actuator
electronics.

\subsection{Actuator}\label{sec:actuator}

The actuator is realized with an arbitrary waveform generator
(AWG). When it receives the feedback trigger, the AWG generates
a feedback pulse with a sampling rate of $1\GHz$. In our
experiment, the actuator pulse (AP) has a duration of
$\tauAP=28\ns$ and uses the derivative removal by adiabatic gate
(DRAG) technique~\cite{Motzoi2009,Gambetta2011a} to prevent
transitions to higher-excited states of the transmon outside of
the subspace spanned by the states $\lvert g\rangle$ and $\lvert
e\rangle$. We typically generate the actuator pulse with a
carrier frequency of $\hbox{100--300}\MHz$ limited by the
bandwidth of the AWG and analog mixer. In the experiment
presented in \secref{experiment} we chose a carrier frequency of
$100\MHz$ for the actuator pulse. We use an analog mixer to
up--convert the actuator pulse to the qubit transition
frequency, which is typically in the range of $\hbox{4--6}\GHz$. Forwarding this pulse to the qubit realizes a
conditional quantum gate on the qubit closing the feedback
loop.

\subsection{Latencies}\label{sec:latencies}

We define the latency $\tauFB$ of the feedback loop
[\figref{loop}\panel{a}] as the time from the beginning of the
readout pulse until the completion of the feedback pulse, i.e.
\begin{equation}\label{eq:tauFB}
    \tauFB \equiv \tauELtot + \tauRO + \tauAP,
\end{equation}
where $\tauELtot$ is the total electronic delay of the signal in
the analog and digital components and cables of the feedback
loop, $\tauRO$ the readout duration (see~\secref{readoutScheme})
and $\tauAP=28\ns$ is the length of the actuator pulse
(see~\secref{actuator}).
We measured the total electronic delay \markred{$\tauELtot =
(219\pm2)\ns$} in-situ by changing the up--conversion frequency
of the feedback pulse to the resonance frequency of the readout
resonator and adjusting the amplitude of the pulse. The resonant
feedback pulse is transmitted through the resonator which makes
it possible to determine the timing of the feedback pulse
relative to the readout pulse. By adding up the contributions
according to \eqref{tauFB} we infer a feedback latency of
\markred{$\tauFB = (352\pm3)\ns$}.

The electronic delay
\begin{equation}
\tauELtot \equiv \tauProc + \tauADCDIO + \tauAWG + \tauGtot,
\end{equation}
can be broken up into accumulated contributions. The signal
processing, which we implemented in the FPGA, introduces a
processing delay of three clock cycles ${\tauProc=30\ns}$~(see
\secref{digital}). The feedback trigger is delayed by
\markred{${\tauProc+\tauADCDIO=(110\pm3)\ns}$} with respect to
the analog input signal, where $\tauADCDIO$ is the delay
introduced by the ADC and digital interfaces~(see
\appref{latencyADCDIO}).

By subtracting the separately determined quantities $\tauProc$,
$\tauADCDIO$ and $\tauAWG$ from the total electronic delay
$\tauELtot$ we estimate the inferred total group delay
\markred{${\tauGtot = (69\pm7)\ns}$} in the cables and analog
components. We expect the total cable length connecting the
analog and digital components to be the dominant contribution to
the inferred group delay. The inferred group delay corresponds
to an approximate total cable length of \markred{$14\m$}
considering an effective dielectric constant
$\epsilon_\mathrm{eff}\approx2$ for the coaxial cables with PTFE
dielectric. This inferred total cable length is consistent with
the experimental setup. The cable length in our setup could be
reduced further by placing the individual components of the
feedback loop closer to each other which can be achieved, for example, by placing the
FPGA and control electronics inside the dilution
refrigerator~\cite{Hornibrook2015,ConwayLamb2016,Homulle2017}.


\section{FPGA-based digital signal processing}
\label{sec:digital}

In this section, we describe our digital signal processing (DSP)
circuit which we implemented on the \hbox{\Virtex4} FPGA. To
derive feedback triggers, the DSP circuit (\figref{firmware})
determines the qubit state by digital demodulation of the
readout signal~(see \secref{overview}). We start by discussing
the digitization and synchronization of the input signal. Next,
we discuss the signal processing features of each block and the
corresponding latencies. Details of the FPGA implementation of
each signal processing block are discussed in
\appref{FPGAimplementation}. We analyze the FPGA timing and
resource usage for the implementation of the DSP circuit on the
Xilinx \hbox{\Virtex4}, \hbox{\Virtex6} and \hbox{\Virtex7} FPGA
in \appref{FPGAtimingResources}.

\subsection{Digitization of the input signal}\label{sec:ADC}

\begin{figure}
    \includegraphics{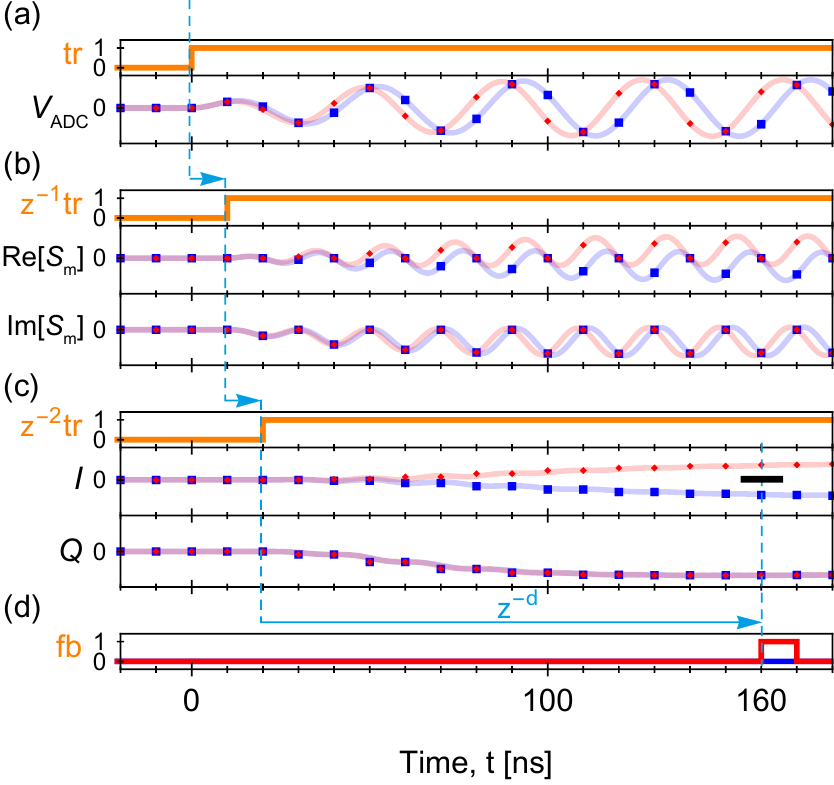}
\caption{Calculated signals at different processing stages for
exemplary inputs when the qubit is either in the ground state
(blue line) or in the excited state (red line). Blue squares and
red diamonds represent the corresponding simulated digital
signals obtained from a simulation of the FPGA design. The
vertical axes have arbitrary units. The blue arrows and dashed
lines visualize the delays of the signals relative to each
other.
(a) The signals $S_\mathrm{ADC}$ from the ADC with two different
phases depending on the qubit being in the ground (blue) or
excited state (red) together with the corresponding trigger
($\tr$) signals (orange line).
(b) Real ($\Re[S_\mathrm{m}]$) and imaginary
($\Im[S_\mathrm{m}]$) part of the complex signal at the output
of the digital mixer with the corresponding trigger delayed by
one clock cycle ($z^{-1}\tr$).
(c) In-phase ($I$) and quadrature ($Q$) component of the signal
obtained at the output of the FIR filter corresponding to a
moving average of four consecutive points with the corresponding
trigger delayed by two clock cycles ($z^{-2}\tr$)
(d) Feedback trigger ($\fb$) conditioned on a threshold on $I$
indicated by the thick horizontal bar at $t = 160\ns$ which is
set by the user-definable delay $z^{-d}$ of $d=14$ clock cycles.
}
    \label{fig:dspTraces}
\end{figure}

Before entering the DSP circuit, the readout signal is digitized
by an external ADC chip which samples the signal with rate
$\fs=100~\mathrm{MS/s}$. Typical readout signals are sine waves
with qubit-state-dependent amplitude and phase as shown in
\figref{dspTraces}\panel{a}. We parameterize the time-dependent
voltage at the input of the ADC as
\begin{equation}\label{eq:VADC}
	\begin{split}
	V_\mathrm{ADC}(t)&= \tilde{A}(t)\cos(\omega_\IF t+\phi(t))\\
&= \frac{\tilde{A}(t)}{2}\left(e^{i(\omega_\IF
t+\phi(t))}+e^{-i(\omega_\IF t+\phi(t))}\right).
	\end{split}
\end{equation}
As discussed in \secref{readoutSchemeImpl}, we choose an
intermediate frequency of ${\omega_\IF/(2\pi)=\fs/4=25\MHz}$ for
the readout signal after analog down--conversion~(see
\secref{readoutSchemeImpl}) which is a useful choice for digital
demodulation as discussed below. The time-dependent amplitude
$\tilde{A}(t)$ is proportional to the amplitude $A(t)$ of the
field at the output of the resonator scaled by the gain of the
analog detection chain and conversion loss of the mixer.

The ADC samples the signal $V_\mathrm{ADC}(t_n)$ at discrete
times $t_n=n/\fs=n\times10\ns$ with index $n$.
The ADC encodes the input voltage range of approximately
$\pm1\V$ as \hbox{14-bit} fixed-point binary values. The
fixed-point representation leads to a discretization step size
of ${2^{-13}\V\approx0.12\mV}$. 
A trigger pulse ($\tr$) is provided together
with the analog signal via a separate digital input of the FPGA
to mark the onset of the readout pulse.

\subsection{Pipelined processing}

\begin{figure}
    \includegraphics{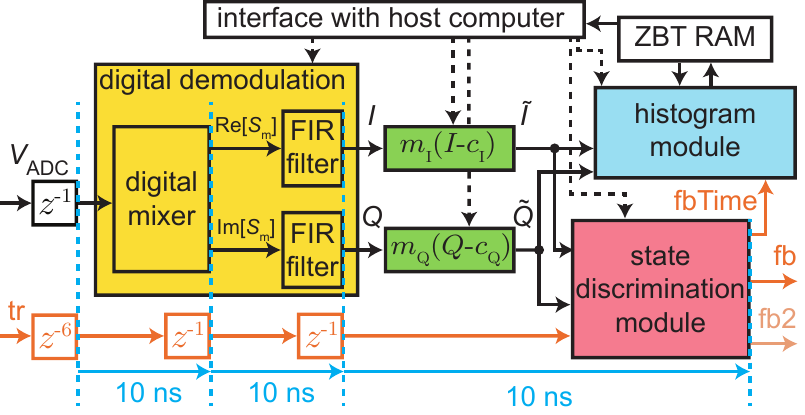}
\caption{Overview of the digital signal processing circuit
showing the flow of the digitized signal (black arrows) and
trigger lines (orange arrows). The symbols $z^{-n}$ denote
delays by $n$ clock cycles implemented with synchronous
D--flip--flops. Blue dashed lines mark positions at which the
signal is further registered in pipelined registers not
explicitly shown. The corresponding latencies of the pipeline
stages are written below the blue arrows. Dotted lines indicate
settings defined via the interface with the host computer.
Explanations of each circuit block are given in the text.}
    \label{fig:firmware}
\end{figure}

We designed the DSP circuit to process the signal from the ADC
in a pipelined manner. The signal from the ADC is initially
buffered in a register implemented by synchronous D--flip--flops
(ADC $z^{-1}$ block in \figref{firmware}) which forward the
value of the signal at each event of a rising edge of the
sampling clock to the next processing element in the pipeline.

A separate trigger input ($\tr$, orange lines in
\figref{firmware}) marks the beginning of each experimental
repetition. In order to synchronize the trigger with the ADC
signal, the trigger initially goes through six pipelined
registers ($z^{-6}$ in \figref{firmware}), which compensate the
difference in delay between the ADC line and trigger line. To
synchronize the signal processing with the sampling clock, we
insert further pipelined registers into the signal and trigger
lines at specific points in the circuit (blue dashed lines in
\figref{firmware}).

\subsection{Digital demodulation}\label{sec:DDC}

As discussed in~\secref{readoutSchemeImpl}, we digitally
demodulate the readout signal to obtain the $I$ and $Q$
components of the signal. Digital demodulation is achieved by
digital frequency down--conversion which involves digital mixing
of the signal with a digital reference oscillator followed by
digital low--pass filtering to remove noise and unwanted
sideband frequency components~\cite{Lyons2011}. 

\subsubsection{Digital mixing}

In the first part of the digital demodulation circuit (yellow
box in \figref{firmware}), we implement a digital mixing
method~\cite{Considine1983,Lyons2011} (digital mixer in
\figref{firmware}) to obtain a sideband at zero frequency. In
the digital mixer, the input signal $V_\mathrm{ADC}$ as defined
in \eqref{VADC}, is multiplied with a complex exponential with
down--conversion frequency $\omega_\IF$ to obtain a complex
output signal $S_\mathrm{m}$,
\begin{equation}\label{eq:Sm}
	\begin{split}
S_\mathrm{m}(t_n) &\equiv V_\mathrm{ADC}\times e^{-i \omega_\IF
t_n} \\
&= \frac{\tilde{A}(t)}{2}\left(e^{i\phi(t)}+e^{-i(2\omega_\IF
t+\phi(t))}\right).
	\end{split}
\end{equation}
The action of the multiplication is to generate two sidebands
corresponding to the two complex exponentials in \eqref{Sm}; one
is corresponding to the complex signal
${I+iQ\equiv\tilde{A}(t)e^{i\phi(t)}/2}$ and the other leads to
oscillations with frequency $2\omega_\IF$ of the output signals
of the mixer~[\figref{dspTraces}\panel{b}]. The complex signal
${I+iQ}$ is the basis on which we determine the state of the qubit
after filtering out the oscillating sideband~(see following
sections).

In practice, the real ($\Re[S_\mathrm{m}]$) and imaginary
($\Im[S_\mathrm{m}]$) parts of the output signal of the mixer
are computed separately by multiplying the input signal with a
discrete cosine to obtain the real part and with a discrete
negative sine to obtain the imaginary part.
The FPGA implementation of the digital mixer is described in
\appref{mixerImplementation}. For $\omega_\IF/(2\pi)=\fs/4$, the
digital mixer introduces a latency of less than one clock cycle
($10\ns$) due to its multiplier-less
implementation~\cite{Considine1983,Lyons2011}. Since the output
signal of the mixer is registered by synchronous D--flip--flops,
the effective latency is one clock cycle. For synchronization,
the trigger signal ($\tr$) is delayed by one clock cycle
[$z^{-1}\tr$ in \figref{dspTraces}\panel{b}].

\subsubsection{Digital low-pass filter}

The second essential part of the digital down--conversion
circuit is a digital low-pass filter, which extracts the $I$ and
$Q$ components from the signals $\Re[S_\mathrm{m}]$ and
$\Im[S_\mathrm{m}]$ by removing the sideband spectral components
oscillating at frequency $2\omega_\IF$~\cite{Lyons2011}. We
implement the digital low-pass filter as a finite impulse
response (FIR) filter~\cite{Lyons2011} which is a discrete
convolution of the digital signal with a finite sequence of
filter coefficients. By matching the filter coefficients
(integration weights) to the expected resonator response, it is
possible to optimize the single-shot readout
fidelity~\cite{Gambetta2007,Johnson2011a,Jeffrey2014,deLange2014,Ng2014,Walter2017}.
While our DSP circuit in principle allows for 40-point FIR
filters with arbitrary filter coefficients, a moving average is
the simplest type of FIR low-pass filter which is possible to
implement without multipliers and therefore has a reduced
processing latency and uses less FPGA resources than a more
general FIR filter. The FPGA implementation of the moving
average module is described in
\appref{movingAverageImplementation}.

The moving average (FIR filter in \figref{firmware}) is applied
separately to the real part ($\Re[S_\mathrm{m}]$) and imaginary
part ($\Im[S_\mathrm{m}]$) of the complex output signal of the
digital mixer, $S_m$, leading to
\begin{equation}
I(t_n) + iQ(t_n) \equiv \frac{1}{l}\sum\limits_{k=n-l+1}^{n}
S_\mathrm{m}(t_k),
\end{equation}
which is a discrete convolution with a square window of length
$l$. In the limit of negligible modulation bandwidth,
the moving average filters a sinusoidal perfectly if the
window length $l$ is a multiple of the oscillation period. In
the case of $\omega_\IF/(2\pi) = \fs/4$, the periodicity of the
unwanted terms at $2\omega_\IF$ is equal to two discrete
samples. Therefore any window length which spans an even
number of samples is suitable to filter out the $2\omega_\IF$
sideband.

The output of the moving average with window length $l=4$ is
shown in \figref{dspTraces}\panel{c}. The $I$ and $Q$ signals at
the output of the moving average show a smooth ramp towards a
steady-state value. In the simulated signals shown in
\figref{dspTraces} an appropriate global phase
offset has been chosen such that the difference between
the traces corresponding to the $\lvert g\rangle$ and $\lvert
e\rangle$ state is maximized in the $I$ component of the
signal~(see \secref{readoutScheme}).

The moving average module has a latency of one clock cycle. The
trigger is delayed accordingly by one additional clock cycle
($z^{-1}z^{-1}\tr = z^{-2}\tr$) for synchronization.

\subsection{Offset subtraction and scaling}

Following the FIR filter block, the $I$ and $Q$ signals enter
blocks which perform offset subtraction and scaling of the
signal (green boxes in \figref{firmware}). The main purpose of
offset subtraction is to set a threshold value as
described in \secref{stateDeterminationModule}. Moreover,
offset subtraction and scaling allows to make best use of the
fixed range and resolution used for recording histograms (see
\secref{histogramModule}).

The outputs of the offset subtraction and scaling blocks are
described by \begin{align} \tilde{I}(t_n) &\equiv
m_\mathrm{I}(I(t_n)- c_\mathrm{I})\\ \tilde{Q}(t_n) &\equiv
m_\mathrm{Q}(Q(t_n) - c_\mathrm{Q}), \end{align} where $c_I$ and
$c_Q$ are offsets in the I/Q plane and $m_I$ and $m_Q$ are
multiplication factors. We determine the parameters $(c_I,c_Q)$ and $(m_I,m_Q)$ in a
calibration measurement. The latencies of the offset subtraction
and scaling blocks are less than one clock cycle and no
synchronous D--flip--flops are used.

\subsection{State discrimination module}
\label{sec:stateDeterminationModule}

The state discrimination module (red box in \figref{firmware})
determines the state of the qubit based on the preprocessed
input signals $\tilde{I}$ and $\tilde{Q}$. Due to the offset
subtraction, the threshold value for state discrimination can be
kept fixed at zero which simplifies the FPGA implementation of
the state discrimination module as discussed in
\appref{stateDeterminationModuleImplementation}.

The readout time $\tauRO$ relative to the onset of the readout
pulse~(see \secref{readoutScheme}) is specified with a variable
delay of $d$ clock cycles after the detection of the trigger
signal, i.e. $d\times10\ns = \tauRO$. In the example shown in
\figref{dspTraces}\panel{c}, the $\lvert g\rangle$ and $\lvert
e\rangle$ states of the qubit are discriminated based on a
threshold value (thick horizontal bar) defined for the $I$
signal at a time $t=160\ns$ which is $d=14$ clock cycles after
the detection of the trigger signal $z^{-2}\tr$. The simulated
$I$ signals corresponding to the $\lvert 0\rangle$ [blue curve
in \figref{dspTraces}\panel{c}] and $\lvert 1\rangle$ [red curve
in \figref{dspTraces}\panel{c}] state are well distinguishable
at the time when the threshold is checked, such that the state
of the qubit can be determined successfully even in presence of
noise~(see \secref{experiment}). The state discrimination module
either issues the feedback trigger [red curve in
\figref{dspTraces}\panel{d}] or does not issue the feedback
trigger [blue curve in \figref{dspTraces}\panel{d}] based on the
determined qubit state.

Our DSP circuit provides the possibility to derive a second
feedback trigger ($\fb2$ in \figref{firmware}) based on both the
in-phase ($\tilde{I}$) or quadrature ($\tilde{Q}$) signal
components. For example, in the quantum teleportation
protocol~\cite{Bennett1993} the states of two qubits at the
sender's location are measured in order to perform a
state-dependent rotation on a qubit at the receiver's location.
In our experimental realization of the teleportation protocol as
discussed in \citeref{Steffen2013}, we discriminated the states
of the two sender's qubits based on two threshold values defined
for the $I$ and $Q$ signals. Based on the outcome of comparing
the $I$ and $Q$ signals to the two threshold values, we issued
two independent trigger signals to two separate AWGs in order to
implement a conditional operation on the receiver's
qubit~\cite{Steffen2013}.

\subsection{Histogram module}\label{sec:histogramModule}

The histogram module records how often the values of the signals
$\tilde{I}$ and $\tilde{Q}$ obtained from a specific integration
window fall into a particular histogram bin when the experiment
is repeated many times. The bins are defined by subdividing the
signal range from -1 to +1 into typically 128 bins. From the
histogram, an estimate of the probability density function of
the signal at the specified times is obtained.

We typically repeat the experiment $10^5$--$10^7$ times to
obtain standard deviations of less than a part per thousand for
the counts in each histogram bin. Storing the histogram of the
signal needs less memory than storing the value of the signal in
each repetition if the number of repetitions exceeds the number
of histogram bins. The histogram module therefore allows for
data reduction at the time when the data is recorded.

We have used the histogram module in previous experiments to
characterize the quantum statistics of microwave radiation emitted from
circuit QED
systems~\cite{Eichler2011a,Eichler2011,Eichler2012,Eichler2014a}.
In the context of feedback experiments, we record histograms to
obtain the probabilities of observing a particular qubit state
in two consecutive qubit readouts as described in
\secref{experiment}.

We update the histogram at the same time as the state
discrimination module determines the qubit state in order to
analyze the readout fidelity and feedback
performance~(see~\secref{experiment}). We synchronize the state
discrimination module and the histogram module using a marker
signal ($\fbTime$ in \figref{firmware}) which is sent from the
state discrimination module to the histogram module. We use an
external Zero Bus Turnaround (ZBT) Random Access Memory
(RAM)~(see \figref{firmware}) to store the histogram. When the
recording of the histogram is completed, we transfer the
histogram to the host computer via the interface. The
implementation details of the histogram module are described in
\appref{histogramModuleImplementation}.


\section{Qubit state initialization experiment}
\label{sec:experiment}

In this section, the functionality of the presented DSP circuit
is demonstrated in the context of a qubit state initialization
experiment. In the experiment we use the feedback loop to reset
the state of a superconducting
qubit~\cite{Reed2010,Riste2012,Riste2012b,Geerlings2013}~(see
\appref{experimentalParameters}) deterministically into its
ground state, independent of its initial state. We correlate the
outcomes of two consecutive qubit measurements in order to
separate out the different effects such as the qubit lifetime
and readout fidelity which contribute to the overall performance
of the feedback protocol.

We choose the repetition period $10\us$ of the experiment to be
longer than the qubit lifetime $T_1\approx1.4\us$, such that the
qubit is approximately in thermal equilibrium with its
environment at the beginning of each experimental repetition. We
observe a finite thermal population
\markred{$\Ptherm\approx7\%$} of the excited state $\e$ due to
the elevated effective temperature of about \markred{$114\mK$}
of the system on which the experiments were performed~(see
\appref{thermal}).

In order to test the feedback protocol, we prepare an equal
superposition of the computational states $\lvert g\rangle$ and
$\lvert e\rangle$ of the superconducting qubit. This choice of
initial state will ideally lead to equal probabilities to find
the states $\lvert g\rangle$ and $\lvert e\rangle$ when the
qubit is measured. Preparing an equal superposition as an
initial state will therefore test the feedback actuator for both
computational states $\g$ and $\e$ of the qubit. An additional
data set (\appref{thermal}) shows that the feedback scheme can
also be used to reduce the thermal population of the excited
state~\cite{Johnson2012,Riste2012,Riste2012b}, providing an
additional benchmark for our feedback loop.

\begin{figure*}
    \includegraphics{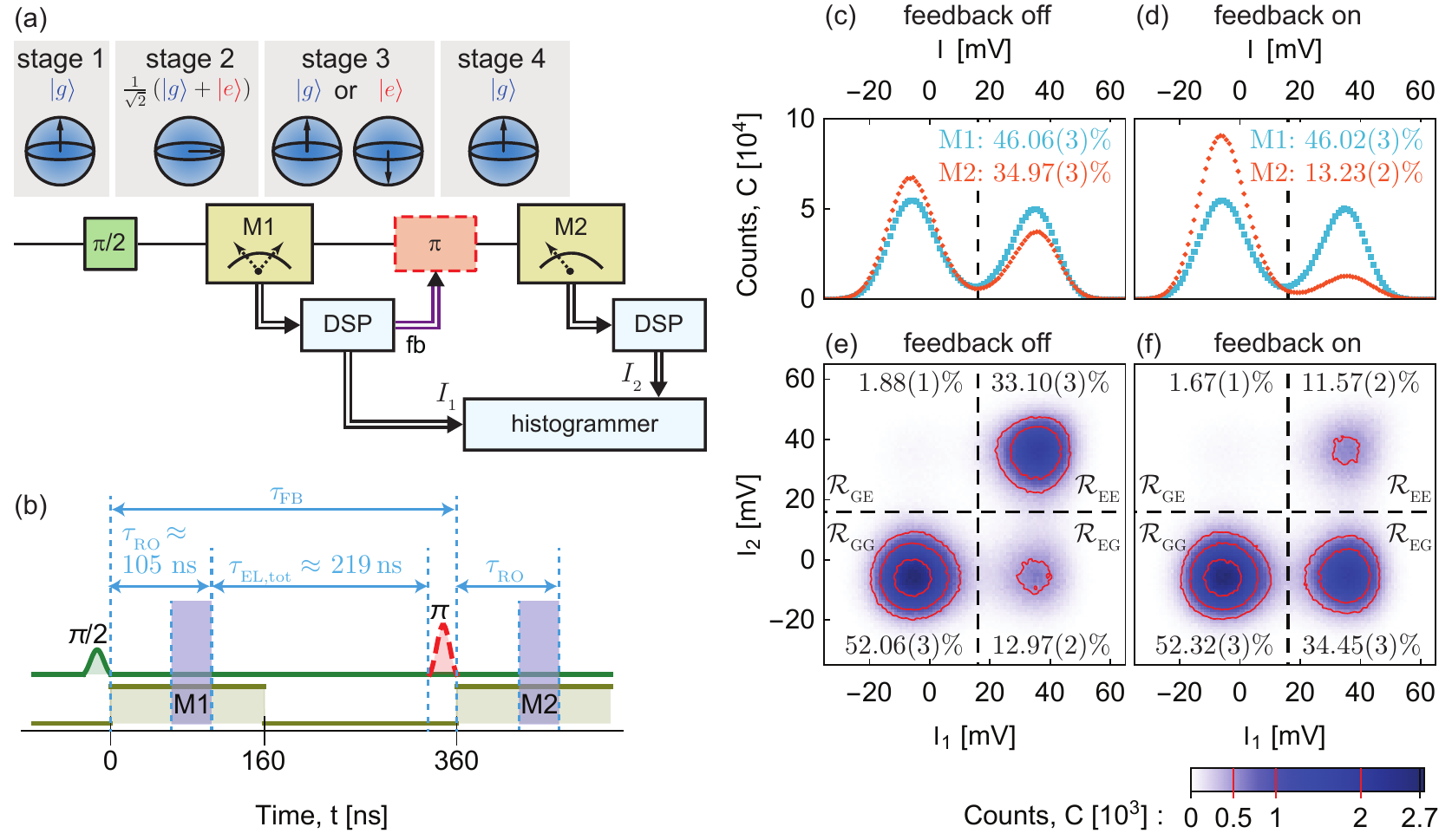}
    \caption{
(a) Quantum circuit depicting the experimental protocol to test
the feedback routine. The state at each stage of the protocol is
represented on the Bloch sphere (blue). The horizontal black line
indicates the evolution of the qubit state over time. Double
arrows ($\Rightarrow$) represent the flow of classical
information. The sequence of operations is: a $\pi/2$ rotation
(green box) of the Bloch vector about an equatorial axis, a
first projective measurement ($\Mone$), a conditional $\pi$
rotation (red dashed box) that depends on the feedback trigger
($\fb$) determined by the digital signal processing (DSP) and a
second projective measurement ($\Mtwo$) of the qubit state.
(b) Pulse scheme showing the timing of microwave pulses applied
to the qubit (green trace), the pulses applied to the resonator
(yellow trace) and the conditional $\pi$ pulse applied to the
qubit (red, dashed). The blue shaded regions mark the
integration windows of the measurements $\Mone$ and $\Mtwo$. The
time offset $\tauRO$ marks the time from the beginning of each
readout pulse to the end of the corresponding integration
window, $\tauELtot$ (blue arrow) marks the delay in the feedback
electronics and $\tauFB$ marks the total feedback latency as
defined in the main text.
(c) Histograms of the in-phase signal $I_1$ obtained from the
first readout pulse M1 (blue dots) and in-phase signal $I_2$
obtained from the second readout pulse M2 (orange dots) for the
case when the feedback actuator is disabled. The dashed
line marks the feedback threshold. For M1 and M2 the percentage
of counts on the right side of the threshold is indicated.
(d) The same type of histograms as in (c) but with the feedback
actuator enabled.
(e) Two-dimensional histogram with $128\times128$ bins counting
the combined outcomes of the first readout $I_1$ (horizontal
axis) and second readout $I_2$ (vertical axis) for the case when
the feedback actuator is disabled. The plane is divided into
four regions \hbox{($\RGG$, $\RGE$, $\REG$, $\REE$)} separated
by the threshold (dashed lines). The percentage of counts
relative to the total count is indicated in each quadrant. Red
lines are contour lines marking specific counts of $\{0.5, 1,
2\}\times 10^3$. (f) The same type of two-dimensional histogram
as in (e) but with the feedback actuator enabled.}
    \label{fig:resultsCombined}
\end{figure*}

Ideally, we consider the case when the qubit is initialized in
the state $\g$ corresponding to the Bloch vector pointing to the
upper pole of the Bloch sphere [stage~1 in
\figref{resultsCombined}\panel{a}]. A microwave pulse at
frequency $\omega_\mathrm{q}$ [green line in
\figref{resultsCombined}\panel{b}] is applied to the qubit to
realize a $\pi/2$ rotation which brings the qubit into the
superposition state ${\lvert+\rangle \equiv (\lvert
g\rangle+\lvert e\rangle)/\sqrt{2}}$ corresponding to a Bloch
vector pointing at the equator of the Bloch sphere [stage~2 in
\figref{resultsCombined}\panel{a}].

When the qubit initially is in state $\e$, for example due to
the non-zero temperature of the system, the effect of the
$\pi/2$ rotation is to prepare the state ${\lvert-\rangle
\equiv (\lvert g\rangle-\lvert e\rangle)/\sqrt{2}}$ which is an
equal superposition of $\g$ and $\e$ with a different phase. The
states $\lvert+\rangle$ and $\lvert-\rangle$ are
expected to lead to an identical distribution of outcomes in the
state detection.

In the experiment, directly after the preparation of the initial
state, at time $t_\Mone=0$, the state of the qubit is measured
with a readout pulse of length $160\ns$~(see $\Mone$ in
\figref{resultsCombined}) applied to the resonator. The
dispersive readout projects the state of the qubit into either
the ground or excited state corresponding to the upper and lower
pole of the Bloch sphere [stage~3 in
\figref{resultsCombined}\panel{a}]. The DSP~(see
\secref{digital}) extracts the in-phase component $I_1$ during
the readout pulse $\Mone$. We filter the signal $I_1$ with a
moving average of four consecutive samples, corresponding to an
integration window [blue region M1 in
\figref{resultsCombined}\panel{b}] of $40\ns$.
We extracted the time $\tauRO\approx105\ns$ of the end of the
integration window~\footnote{It is
in principle possible to shorten the $160\ns$ duration of the
readout pulse to match the end of the integration window at
$t=105\ns$ but we keep the length of the readout pulse constant
to simplify the calibration procedure.} relative to the beginning of the readout
pulse by fitting a theoretical model to the switch-on dynamics
of the readout signal in a time-resolved measurement~\cite{Walter2017}.

The histogram of $I_1$ [blue
dots in \figref{resultsCombined}\panel{c}] reveals two Gaussian
peaks corresponding to the distributions of the in-phase signal
for the qubit being in state $\g$ or $\e$. The measured initial
excited state probability ${\Prob[E_1]_\text{fb off} =
46.06(3)\%}$, is the fraction of counts of values $I_1$ above
the threshold value \markred{${I_\mathrm{t} = 16\mV}$} [dashed
line in \figref{resultsCombined}\panel{c}] relative to the total
count $C_\mathrm{tot} = 2'097'152$ of measurements.

With a master equation~\cite{Carmichael2002} we simulate the
decay of the qubit state with characteristic time $T_1=1.4\us$
during the time of the $\pi/2$ pulse and the readout up to the
center of the integration window
[see~\figref{resultsCombined}\panel{b}]. Furthermore we take
into account a bias of the measured probabilities towards $50\%$
due to the finite readout error of~$3\%$~(see \appref{readout}).
From the master equation simulation we obtain an expected
excited state probability of \markred{${\Prob[E_1]_\text{sim} =
47.07\%}$} in the first measurement $\Mone$ which agrees
reasonably with the measured probability ${\Prob[E_1]_\text{fb
off}}$ (see above). A source of systematic errors is
measurement-induced mixing~\cite{Slichter2012}. An additional
reason for the systematic deviation of the measured probability
from the simulated probability is that the chosen threshold
value \markred{$I_\mathrm{t} = 16\mV$} deviates from the value
\markred{$I_\text{t, opt}\approx13\mV$} which optimizes readout
fidelity~(see \appref{readout}). This offset leads to a bias of
the observed probabilities towards the ground state in addition
to a systematic bias due to state transitions during the
integration time~\cite{Walter2017}.

The feedback loop is configured to deterministically prepare the
state $\g$ [stage~4 in \figref{resultsCombined}\panel{a}]. The
feedback pulse, inducing a $\pi$~rotation of the Bloch vector of
the qubit, turns the state $\e$ into $\g$ and vice versa. Thus,
the feedback $\pi$~pulse is issued only if the first measurement
$\Mone$ revealed the qubit to be in state $\e$. The $\pi$ pulse
[red dashed line in \figref{resultsCombined}\panel{b}] arrives
at the qubit with delay of $\tauELtot$~(see \secref{latencies})
conditioned on the readout result of $\Mone$.

For verification, a second readout pulse ($\Mtwo$ in
\figref{resultsCombined}) is applied to the qubit at the time
$t_\Mtwo=360\ns$ directly after the arrival of the feedback
pulse at the qubit. The difference between $t_\Mtwo$ and the
beginning of the first readout pulse corresponds to the
total feedback latency $\tauFB$~(see \secref{latencies}). We
recorded histograms of $I_2$, which is the filtered in-phase
component of the signal at time $t_\Mtwo+\tauRO$. When the
feedback actuator is disabled, the histogram of $I_2$ [orange
dots in \figref{resultsCombined}\panel{c}] shows reduced counts
on the right side of the threshold with an excited state
probability of ${\Prob[E_2]_\text{fb off} = 34.97(3)\%}$.
Extending the master equation simulation introduced above to
include the full pulse sequence up to the second readout pulse,
we obtain \markred{${\Prob[E_2]_\text{fb off, sim} = 37.89\%}$}
in reasonably good agreement with the measured value. The state
decay between $\Mone$ and $\Mtwo$, which leads to the observed
reduction in the excited state population, causes errors in the
feedback action as discussed below.

When the experiment is repeated with the feedback actuator enabled, the
double-peaked histogram obtained from the first readout $I_1$
[blue dots in \figref{resultsCombined}\panel{d}] is approximately
identical to the case without feedback, as expected, with the
measured excited state probability ${\Prob[E_1]_\text{fb on} =
46.02(3)\%}$ agreeing with ${\Prob[E_1]_\text{fb off}}$ within
the statistical error bars. After the feedback pulse, 
 in the histogram of $I_2$ [orange dots in
\figref{resultsCombined}\panel{d}], the measured excited state
probability is significantly reduced to ${\Prob[E_2]_\text{fb
on}=13.23(2)\%}$. This probability compares reasonably well with
the simulated value of \markred{${\Prob[E_2]_\text{fb on, sim} =
10.50\%}$} obtained from the master equation simulation
introduced above. We attribute the difference between the measured
and simulated value of ${\Prob[E_2]_\text{fb on}}$ to
measurement-induced mixing and the deviation of the
feedback threshold from the optimal value (see above).

To obtain a figure of merit for the feedback protocol that is
independent of characteristics such as state decay and
temperature of the quantum system, we study correlations between
the outcomes of the two readout pulses $\Mone$ and $\Mtwo$. From the
two-dimensional histograms [\figref{resultsCombined}\panel{e,f}]
with axes $I_1$ and $I_2$, we obtain experimental probabilities
to observe a specific range $\mathcal{R}$ of two consecutive
measurement outcomes $(I_1,I_2)$. The probabilities
${\Prob[\mathcal{R}_{xy}]}$ correspond to observing the qubit in
state~$x$ with the first readout pulse and consecutively in
state~$y$ with the second readout pulse. These probabilities are
obtained from the normalized counts in the four quadrants
\hbox{($\RGG$, $\RGE$, $\REG$, $\REE$)} separated by the
threshold [dashed lines in \figref{resultsCombined}\panel{e,f}].

When the feedback is enabled, the measured probability
${\Prob[\REE]_\text{fb on} = 11.57(2)\%}$
[\figref{resultsCombined}\panel{f}] corresponds to the unwanted
event of the state $\e$ being observed consecutively with both
readout pulses. We explain the dominant contribution to
${\Prob[\REE]_\text{fb on}}$ by state decay between the first
readout pulse and the feedback pulse. The probability of
state decay between the first and second readout pulse is
extracted from a reference measurement of ${\Prob[\REG]_\text{fb
off} = 12.97(2)\%}$ [\figref{resultsCombined}\panel{e}] when the
feedback is disabled. The probabilities ${\Prob[\REE]_\text{fb
on}}$ and ${\Prob[\REG]_\text{fb off}}$ are close to each other
since the conditional $\pi$ pulse swaps the state $\g$ with $\e$
before the second readout pulse. The corresponding simulated
probabilities ${\Prob[\REE]_\text{fb on, sim} = 8.32\%}$ and
${\Prob[\REG]_\text{fb off, sim} = 11.37\%}$~(see
\tabref{jointP}) are in reasonable agreement with the
experimental values considering the sources of systematic errors
as discussed above.

The measured probability
${\Prob[\RGE]_\text{fb on} = 1.67(1)\%}$
[\figref{resultsCombined}\panel{f}] of a transition from state
$\g$ to $\e$ when the feedback loop is enabled is close to the
reference value ${\Prob[\RGE]_\text{fb off} = 1.88(1)\%}$
[\figref{resultsCombined}\panel{e}] when the feedback is
disabled. This shows that the state is correctly left unchanged
when the qubit is already in state $\g$. A possible reason for
the small systematic deviation of ${\Prob[\RGE]_\text{fb on}}$
from ${\Prob[\RGE]_\text{fb off}}$, which is on the order of
$0.2\%$, could be drifts in the experimental parameters such as
the phase of the readout signal.

In summary, the probabilities of the combined events
(\tabref{jointP}) show that in the feedback protocol
the $\pi$ pulse is applied only when it is intended and that the
probability of the unwanted events in region $\REE$ is limited
by state decay between the first measurement and the feedback
pulse.

\begin{table}
\caption{\label{tab:jointP} Experimental (exp.) and simulated
(sim.) probabilities ${\Prob[\mathcal{R}_{xy}]}$ of the events
to observe the qubit in state~$x$ in the first measurement and
in state~$y$ in the second measurement when the feedback is
either disabled (off) or enabled (on). The simulated values are
obtained from a master equation simulation. See main text for
details.}
\begin{ruledtabular}
\begin{tabular}{l|rr|rr}
	& \multicolumn{2}{c|}{feedback off}
	& \multicolumn{2}{c}{feedback on} \\
	& exp.& sim.  & exp.  & sim. \\
	\hline
	$\Prob[\RGG]$ & 52.06(3)\% & \markred{50.74\%} & 52.32(3)\% &
	\markred{50.74\%} \\
	$\Prob[\RGE]$ & 1.88(1)\% & \markred{2.18\%} & 1.67(1)\% &
	\markred{2.18\%} \\
	$\Prob[\REG]$ & 12.97(2)\% & \markred{11.37\%} & 34.45(3)\% &
	\markred{38.75\%} \\
	$\Prob[\REE]$ & 33.10(3)\% & \markred{35.71\%} & 11.57(2)\% &
	\markred{8.32\%} \\
\end{tabular}
\end{ruledtabular}
\end{table}


\section{Conclusions and discussion}

We developed a low-latency FPGA-based digital signal
processing unit for quantum feedback and feedforward
applications such as the qubit initialization scheme presented
in this paper and the deterministic quantum teleportation
realized in~\citeref{Steffen2013}.

Our experimental results show that the feedback loop performs as
expected. The total feedback latency amounts to
\markred{$\tauFB=(352\pm3)\ns$} determined by the sum of ADC
latency, processing latency, AWG latency, cable delays, readout
time and feedback pulse duration. To reduce the probability of
state decay between the state detection and the feedback action,
the ratio $r\equiv\tauFB/T_1$ of the feedback latency to the
qubit lifetime $T_1$ needs to be reduced.
Since the probability of state decay is expected to be
proportional to $1-\exp(-r)$, a $T_1$ time of about $40\us$
would be needed to achieve error probabilities of less than
$1\%$ in one iteration of the feedback scheme presented in this
work. Conversely, with the longest $T_1$ times achievable with
state-of-the art superconducting circuits of up to approximately
$100\us$~\cite{Rigetti2012,Barends2013,Yan2016a}, feedback
latencies of less than $100\ns$ would be needed to reduce the
error probability to less than one part per thousand. In the
present work, we demonstrated digital processing latencies on
the order of $30\ns$, which are among the shortest latencies
reported for FPGA-based signal
analyzers~\cite{Campagne-Ibarcq2013,Riste2013,Ryan2017} in the
context of superconducting qubits. Simultaneously, the usage of
advanced readout strategies enables a shorter optimal
readout times~\cite{Jeffrey2014,Walter2017}. Shorter latencies
for analog--to--digital conversion and cable delays may be
achievable by using custom-made circuit boards which work at
cryogenic
temperatures~\cite{Hornibrook2015,ConwayLamb2016,Homulle2017} or
by on-chip logical
elements~\cite{Yamamoto2014,Andersen2016,Balouchi2017}.

Low latency feedback loops may play a role in realizing future
quantum computers, where a key ingredient is quantum error
correction~\cite{Fowler2012,Lidar2013,Terhal2015n} in which
error syndromes of a quantum error correction code are detected
by repetitive measurements. The syndrome measurements are
designed to keep track of unwanted bit flip and phase errors. In
this context it is essential to have a flexible low latency
classical processing unit to process the error syndromes without
causing additional delay for the quantum processor. A large set
of quantum error correction codes may work with a passive `Pauli
frame' update~\cite{Knill2005}, however, it still remains an
open question~\cite{OBrien2017} whether some level of correction
and qubit reset using active feedback is preferable. Therefore,
having a low latency signal processor with feedback capabilities
as presented in this work, will be instrumental for scaling up
quantum technologies.

\begin{acknowledgments}
The authors would like to thank Deniz Bozyigit for initial
contributions to the FPGA firmware. The authors further
acknowledge useful discussions with Johannes Heinsoo, Sebastian
Krinner, Markus Oppliger and Lars Steffen.

The authors acknowledge financial support by the National Centre
of
Competence in Research Quantum Science and Technology
(NCCR QSIT), a research instrument of the Swiss
National Science Foundation (SNSF), by the Swiss Federal
Department
of Economic Affairs, Education and Research
through the Commission for Technology and Innovation
(CTI), by the Office of
the Director of National Intelligence (ODNI), Intelligence
Advanced Research Projects Activity (IARPA), via the
U.S. Army Research Office grant W911NF-16-1-0071 and
by ETH Zurich. The views and conclusions contained
herein are those of the authors and should not be interpreted
as necessarily representing the official policies or
endorsements, either expressed or implied, of the ODNI,
IARPA, or the U.S. Government. The U.S. Government
is authorized to reproduce and distribute reprints for
Governmental purposes notwithstanding any copyright
annotation thereon.

\end{acknowledgments}


\appendix

\section{Experimental setup}\label{app:setup}

\begin{figure*}
    \includegraphics{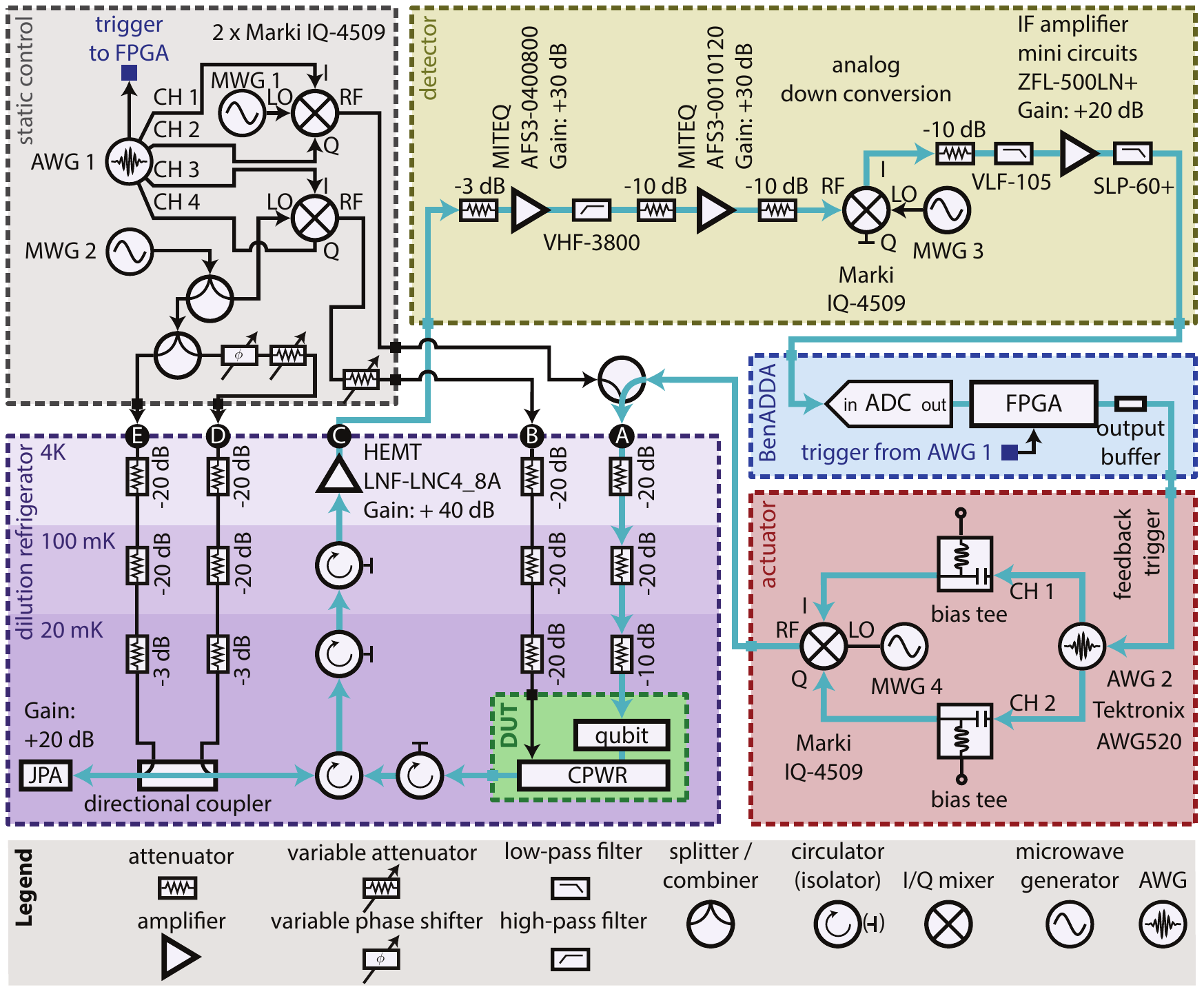}
\caption{Schematic of the experimental setup used for quantum
feedback. Cyan arrows point into the direction of the signal
flow in the feedback loop. The device under test (DUT) is a
superconducting circuit comprised of a qubit coupled to a
coplanar waveguide resonator (CPWR). The color scheme of the
blocks DUT (green), static control (grey), detector (yellow),
Nallatech \BenADDA card (blue) and actuator (red) corresponds
with \figref{loop} in the main text. In addition, the three
different shades of purple in the dilution refrigerator (purple
box) indicate the temperature stages ($20\mK$, $100\mK$ and
$4\K$) to which the corresponding components are thermalized.
The signal ports at the dilution refrigerator are labeled with
letters in circles (A-E).}
    \label{fig:setup}
\end{figure*}

The device under test (DUT, green box in \figref{setup}) is a
superconducting circuit with one superconducting transmon qubit.
The DUT is thermalized to the $20\mK$ stage of a dilution
refrigerator (purple box in \figref{setup}).

Single-qubit quantum gates are realized by driving transitions
between the ground and first excited state of the transmon by
applying microwave pulses through a dedicated microwave line
(port~A in \figref{setup}). The microwave line is thermalized by
attenuators at three temperature stages
$T=(4\K,\,100\mK,\,20\mK)$. The attenuators reduce the signal
and noise coming from the room-temperature electronics and add
Johnson-Nyquist noise at their respective temperature $T$,
thereby reducing the effective temperature of the microwave
radiation in the cable.
The qubit pulses for static control (grey box in \figref{setup})
are generated by AWG 1 and up--converted to microwave
frequencies using an I/Q mixer driven by a local oscillator (LO)
signal from microwave generator MWG~1.

Readout of the qubit is realized by a pulsed measurement of the
transmission of microwaves through a coplanar waveguide
resonator (CPWR). The readout pulse is applied to the CPWR
through the resonator drive line (port~B in \figref{setup}). The
readout pulses are also generated by AWG~1. An I/Q mixer with an
LO signal from MWG~2 allows for shaping the readout pulses which
can be useful to achieve faster ring-up and ring-down of the
intra-cavity field~\cite{Bultink2016,McClure2016}. In order to
adjust the power range for the resonator drive a variable
attenuator is used at the RF output of the mixer.

The transmitted signal is directed through an isolator, circulator, and 
directional coupler to a Josephson parametric amplifier
(JPA)~\cite{Yurke1989} based on a $\lambda/4$ resonator shunted
with an array of SQUID
loops~\cite{Yamamoto2008,Castellanos2008,Eichler2014}. The
isolators and circulators protect the DUT from pump leakage and
thermal noise. The pump tone needed to achieve a gain of
approximately $20\dB$ in the JPA is derived via splitters from
the same microwave generator MWG~2 as is used for the readout
pulses which reduces drifts of relative phase between the two
signals. Low phase noise is essential if the JPA is operated in
a phase-sensitive mode~\cite{Movshovich1990,Eichler2014a}. The
pump signal (port~E in \figref{setup}) is combined with the
signal from the resonator through a directional coupler.

Both the signal and pump tone are reflected from the JPA. To
avoid saturation of the subsequent amplifiers, we destructively
interfere the reflected pump tone with a cancellation tone
applied to the directional coupler (port~D in \figref{setup}).
The phase and amplitude of the cancellation tone are adjusted
using a variable phase shifter and attenuator.

After amplification by the JPA, the signal is passed via
isolators which attenuate reversely propagating radiation
towards a high-electron-mobility transistor (HEMT) amplifier to
further amplify the signal with a gain of $40\dB$ before it
exits the dilution refrigerator (port~C in \figref{setup}).

In the detection electronics (yellow box in \figref{setup}) at
room temperature, the signal is amplified further using
low-noise microwave amplifiers. In order to reduce noise below
the frequencies of interest, the signal is high-pass filtered
with a cut-off frequency of about $4\GHz$. The carrier frequency
of typically $7\GHz$ is converted down to an intermediate
frequency (IF) using an analog I/Q mixer and a separate
microwave generator, MWG~3, for the LO signal. The IF signal at
the $I$ output of the mixer is further amplified using an IF
amplifier and low-pass filters are used to suppress noise
outside the detection bandwidth of the ADC ($50\MHz$).
Attenuators between the amplifiers and the mixer are used to
suppress standing waves due to impedance mismatches and in order
to prevent saturation of the mixer, amplifiers and ADC.

After amplification and analog down--conversion, the signal is
digitized by the ADC and forwarded to the FPGA on the Nallatech
\BenADDA card. The digital signal processing (DSP) circuit which
we implemented on the FPGA generates a feedback trigger
conditioned on the digitized and processed signal~(see
\secref{digital}).

The feedback trigger is forwarded to AWG~2 which is part of the
actuator electronics (red box in \figref{setup}). When receiving
the feedback trigger, AWG~2 generates a pulse which is
up--converted to the qubit frequency, typically at $5$--$6\GHz$,
using an I/Q mixer and LO from microwave generator MWG~4.
Bias--tees allow to compensate unwanted DC offsets of the I/Q
inputs of the mixer in order to suppress LO leakage. The
up--converted microwave pulses are forwarded to the qubit
(port~A in \figref{setup}).

All AWGs, MWGs, as well as ADC and DSP clocks are synchronized
to a $10\MHz$ sine wave from an SRS FS725 rubidium frequency
standard.

\section{Latency of analog to digital conversion and digital input}
\label{app:latencyADCDIO}

The ADC latency and digital input--output latencies of the FPGA
are inferred from the timing relative to the input trigger and
feedback trigger. When the variable delay in the state
discrimination module~(see
\appref{stateDeterminationModuleImplementation}) is set to $d=1$
clock cycle, we measure the delay from the trigger input to the
feedback trigger with an oscilloscope to be $\tau_\text{tr--fb}
= 110\ns\pm3\ns$. Since the input trigger is synchronized with
the digitized signal from the ADC in the DSP circuit, we infer
that the ADC delay and digital input--output delay is $\tauADCDIO
= \tau_\text{tr--fb} - \tau_\mathrm{proc} = 80\ns\pm3\ns$.

The delay $\tauADCDIO$ has several contributions which we did
not determine individually. The pipelined architecture of the
AD6645 ADC introduces a delay of four clock cycles ($40\ns$) and
a latency of one additional clock cycle ($10\ns$) to transfer
the digitized signal from the ADC to the FPGA where it is
registered in a synchronous D--flip--flop. Further delays are
expected to contribute to $\tauADCDIO$ due to the routing of the
digital signal on the \BenADDA board as well as
pad--to--flip--flop and flip--flop--to--pad delays on the
FPGA~(see \appref{FPGAtiming}).

\section{Implementation details of digital signal processing blocks}
\label{app:FPGAimplementation}

Here we specify implementation details of the blocks of the DSP
circuit presented in \secref{digital} which are relevant for the
processing latency.

\subsection{Digital mixer}\label{app:mixerImplementation}

\begin{figure*}
    \includegraphics{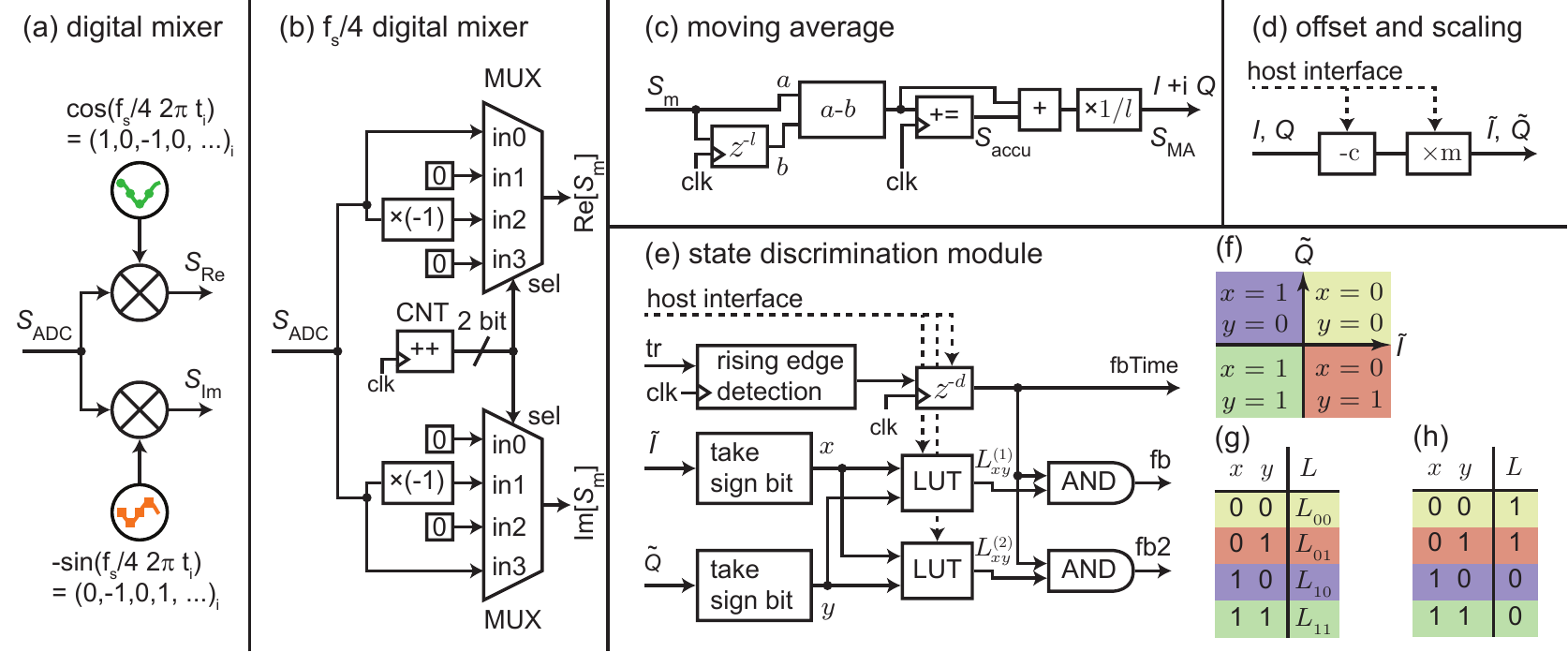}
\caption{Details of the blocks of the DSP circuit relevant for
feedback generation.
(a) Digital I/Q mixer implemented with multipliers (circles with
crosses). (b) Quarter sampling rate $\fs/4$ digital I/Q mixer
implemented with multiplexers (MUX) that forward one of their
inputs to their output based on the selection (sel). The sel
signal is driven by a repeating two-bit counter (CNT,++). (c)
moving average using an accumulator (+=) (d) Circuit for offset
subtraction and scaling.
(e) Schematic of the logic circuit of the state discrimination
module based on lookup tables (LUT). (f) Representation of the
plane spanned by the offset-subtracted in-phase ($\It$) and
quadrature ($\Qt$) signal components. The four quadrants are
labeled by the corresponding values of the sign bits
$x$ and $y$. (g) Representation of one LUT based on the inputs
$x$ and $y$ (first two columns). The
third column contains the symbolic value $L_{xy}$ stored in the
LUT for every combination of
input bits $x$ and $y$. (h) Specific example of how to fill in
the LUT. }
    \label{fig:firmwareDetails}
\end{figure*}

The cosine and sine signals, ${\cos(\omega_\IF t_n)}$ and
${-\sin(\omega_\IF t_n)}$, for digital mixing are typically
generated either using a lookup table with precomputed values or
by an iterative algorithm and then multiplied with two copies of
the signal as shown in \figref{firmwareDetails}\panel{a}. While
these methods work for arbitrary frequencies~$\omega_\IF$, a
simplified method exists for the special case when $\omega_\IF$
equals a quarter of the sampling rate, i.e. ${\omega_\IF/(2\pi)
= \fs/4}$~\cite{Considine1983,Lyons2011}.

In the $\fs/4$ case, the periodic sequences for the cosine and
negative sine are simply ${(1, 0, -1, 0)}$ and ${(0, -1, 0, 1)}$
respectively~\cite{Considine1983}. Since multiplication with
$0$, $1$ and $-1$ is trivial, we replace the multipliers by
counter-driven multiplexers (MUX) that periodically switch
between four inputs as shown in
\figref{firmwareDetails}\panel{b}. The 2-bit repeating counter
(CNT) iterates through a sequence of four values ${(0, 1, 2,
3)}$, jumping to the next value in every clock cycle and
restarting from $0$ after it has reached $3$. The output of the
counter is forwarded to the selection (sel) input of the
multiplexers (MUX). The selection input of the multiplexers
determine which of the four inputs ${(\mathrm{in0},
\mathrm{in1}, \mathrm{in2}, \mathrm{in3})}$ of the multiplexers
are forwarded to their output. The four inputs of the
multiplexer for the real part ($\Re[S_\mathrm{m}]$), correspond to
multiplying the signal with ${(1, 0, -1, 0)}$ while the inputs
of the multiplexer for the imaginary part ($\Im[S_\mathrm{m}]$)
correspond to multiplication with ${(0, -1, 0, 1)}$.

\subsection{Moving average}
\label{app:movingAverageImplementation}

In the following, we discuss how to implement the moving average
(circuit shown in \figref{firmwareDetails}\panel{c}), which is
the simplest type of FIR filter, with a processing latency of
less than one clock cycle ($10\ns$). The moving average is
applied in parallel to the real and imaginary parts of the
output $S_\mathrm{m}$ of the mixer, i.e.~two copies of the
circuit shown in \figref{firmwareDetails}\panel{c} are
implemented with outputs $I$ and $Q$ respectively.

The first step in the circuit for computing the moving average,
as shown in \figref{firmwareDetails}\panel{c}, is to fan out the
input signal into two branches. One branch $b$ is delayed by a
variable delay~($z^{-l}$) of $l$ clock cycles while no operation
is performed on the other branch $a$, i.e. $b_m=a_{m-l}$. A
subtractor then computes the difference $a-b$ between the values
of the two branches which is forwarded to an accumulator [+= in
\figref{firmwareDetails}\panel{c}]. In every clock cycle, the
accumulator adds the value at its input to the sum stored
internally and forwards the updated sum to the output. Therefore
the output $S_\text{accu,n}$ at clock cycle $n$ of the
accumulator is the sum of all input samples up to clock cycle
$n-1$, i.e.
\begin{align}\label{eq:movingAverageAccu}
S_\text{accu,n} &\equiv \sum\limits_{m=0}^{n-1}
(a_m-b_m)
     = \sum\limits_{m=0}^{n-1} (a_m - a_{m-l})\nonumber\\*
     &= \sum\limits_{m=n-l}^{n-1} a_m
+ \underbrace{\sum\limits_{m=0}^{n-l-1}\left( a_{m} -a_{m}
\right)}_{=0}
	- \underbrace{\sum\limits_{m=-l}^{-1}a_{m}}_{=0}\nonumber\\*
     &= \sum\limits_{m=n-l}^{n-1} a_m,&
\end{align}
where the last equality holds assuming that all input samples
with negative index are equal to zero, i.e $a_m=0$ for $m<0$. To
make sure that this assumption holds true, we initialize the
registers of the variable delay and the accumulator to zero. As
depicted in \figref{firmwareDetails}\panel{c}, an additional
adder (+) adds the most recent value of the difference $a_n-b_n$
at the input of the accumulator to its output and a constant
factor of $1/l$ normalizes the moving average. Thus, the final
signal at the output of the moving average module
($S_\text{\text{MA}}$) is
\begin{align}\label{eq:movingAverageFinal}
	S_\text{\text{MA},n}
&\equiv \frac{1}{l}\left(S_\text{accu,n} + a_n -
b_n\right)\nonumber\\*
&=\frac{1}{l}\left(\sum\limits_{m=n-l}^{n-1} a_m + a_n -
a_{n-l}\right)\nonumber\\*
     &=\frac{1}{l}\sum\limits_{m=n-l+1}^{n} a_m.
\end{align}
As opposed to the sums in \eqref{movingAverageAccu}, which stop
at index $n-1$, the final sum in \eqref{movingAverageFinal}
includes the most recent sample with index $n$, which shows that
the additional adder reduces the effective processing latency to
less than one clock cycle.

\subsection{Preprocessing module}
\label{app:preprocessingModuleImplementation}

Offset subtraction ($-c$ in \figref{firmwareDetails}\panel{d})
is implemented with lookup tables (LUTs). The parameter $c$ is
configurable via the interface with the host computer (indicated
by dashed arrows). The multiplication ($\times m$ in
\figref{firmwareDetails}\panel{d}) is implemented without the
use of actual multipliers but rather uses bit shift operations,
which are effective multiplications with powers of two. Avoiding
the allocation of multipliers reduces hardware resource
consumption and leads to reduced processing latencies. The
multiplication is made configurable using multiplexers to choose
between different bit shift operations. The bit shift operation
is chosen via the host computer interface.

\subsection{State discrimination module}
\label{app:stateDeterminationModuleImplementation}

The state discrimination module determines the qubit state and
provides feedback triggers based on the sign bits $x$ and $y$ of
the preprocessed signals $\tilde{I}$ and $\tilde{Q}$ as shown in
\figref{firmwareDetails}\panel{e}. The sign bits of $\tilde{I}$
and $\tilde{Q}$ are $0$ if the respective signal is positive and
$1$ if it is negative, as depicted in
\figref{firmwareDetails}\panel{f}. Due to the prior offset
subtraction, determining the sign bits of $\tilde{I}$ and
$\tilde{Q}$ is equivalent to comparing the $I$ and $Q$ signals
each to an arbitrary threshold value. Two lookup tables (LUT)
define the binary feedback with two independent bits
$L^{(1)}_{xy}$ and $L^{(2)}_{xy}$ which are selected based on
the two sign bits $x$ and $y$ as depicted in
\figref{firmwareDetails}\panel{g}. The entries of the LUT can be
set via the host computer interface [dashed arrows in
\figref{firmwareDetails}\panel{e}]. In the example shown in
\figref{firmwareDetails}\panel{h}, the value of the feedback bit
is $1$ if and only if $x=0$ corresponding to a non-negative
value of the $I$ component of the signal.

The input trigger signal is used as a reference for the timing
of the feedback triggers relative to the onset of the readout
pulse. As shown in \figref{firmwareDetails}\panel{e}, the
trigger first enters a rising edge detection block. The output
of the rising edge detection block is $1$ if and only if the
input binary value of the trigger was $0$ in the previous clock
cycle and $1$ in the present clock cycle. The output of the
rising edge detection is delayed with a variable delay $z^{-d}$
where $d$ is the number of clock cycles (each being $10\ns$)
corresponding to the readout time $\tauRO$, i.e. ${d\times10\ns
= \tauRO}$. The parameter $d$ can be set via the host computer
interface. The output of the variable delay, to which we refer
as the $\fbTime$ marker, marks the specific time at which the
feedback pulse is provided. To assert that the feedback triggers
are issued at the correct time, the feedback triggers $\fb$ and
$\fb2$ are based on the AND operation of the output of the LUT
and the $\fbTime$ marker.

\subsection{Histogram module}
\label{app:histogramModuleImplementation}

\begin{figure}
    \includegraphics{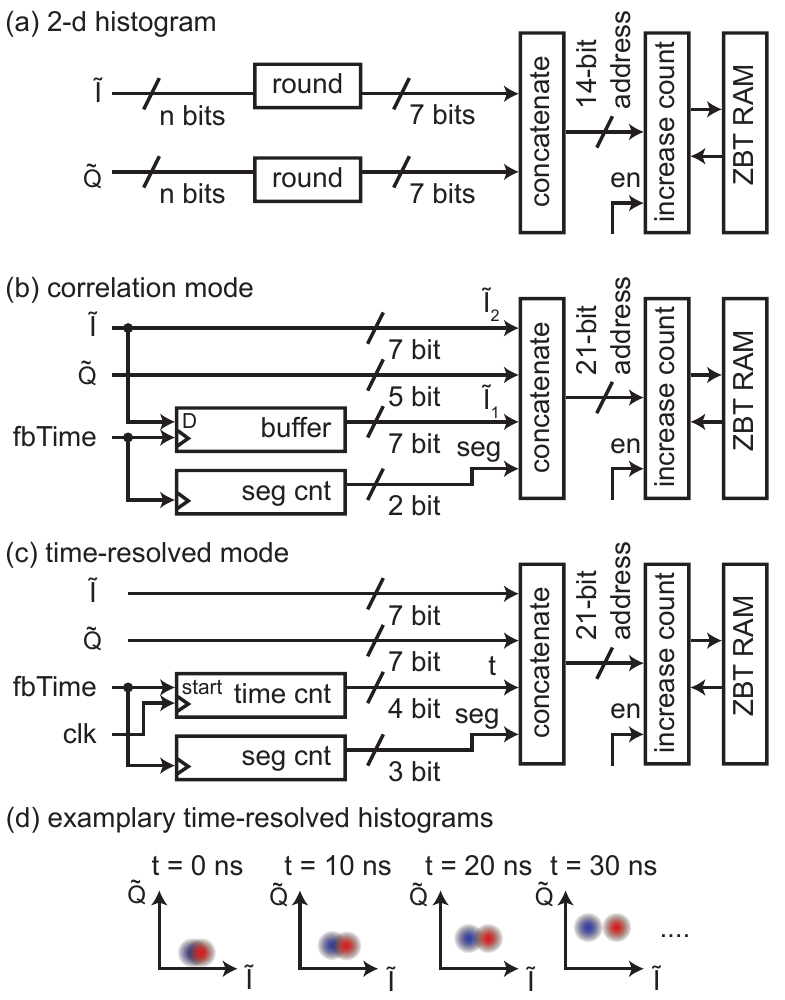}
\caption{(a) Example circuit for recording two-dimensional
histograms with dimensions being the preprocessed 7-bit signals
$\tilde{I}$ and $\tilde{Q}$. See text for details. (b)
Correlation-mode with buffer and segment counter (seg cnt). For
simplicity, the rounding steps are not shown here. (c)
Time-resolved mode with a time counter (time cnt). For
simplicity, the rounding steps are not shown here. (d)
Illustration of exemplary time-resolved histograms of the
$\tilde{I}$ and $\tilde{Q}$ values at four consecutive times $t$
when the qubit state is $\lvert g\rangle$ (blue) or $\lvert
e\rangle$ (red).}
    \label{fig:histogramModule}
\end{figure}

The histogram module is important to assess the feedback
performance and to calibrate the experimental setup. Here we
explain how our multi-dimensional histogram module is
implemented. The histogram module has different operational
modes. We first introduce the circuit for recording
two-dimensional histograms as shown in
\figref{histogramModule}\panel{a}. The input signals $\tilde{I}$
and $\tilde{Q}$ are rounded to 7-bit fixed point numbers which
means that the full range of $\pm1V$ is subdivided into
$2^7=128$ bins. The 7-bit fixed-point representations of
$\tilde{I}$ and $\tilde{Q}$ are concatenated into a 14-bit
address of the histogram bin which stores the number of
occurrences of the combination of values
$(\tilde{I},\tilde{Q})$. The ``increase count'' block manages
the communication with the ZBT RAM in order to increase the
stored count whenever the enable flag (en) is active. For
feedback experiments, the enable flag is derived from the
$\fbTime$ marker such that the histogram is updated when the
feedback decision is made~(see \secref{digital}).

In the correlation mode of our histogram module, a buffer
[\figref{histogramModule}\panel{b}] stores the value of
$\tilde{I}$ at every reception of the $\fbTime$ marker. The
buffered signal $\tilde{I}_1$ is combined with the most recent
signal $\tilde{I}_2$ to record the probability to observe a
specific combination $\tilde{I}_1$ and $\tilde{I}_2$ in two
consecutive readouts~(see \secref{experiment}). In addition a
segment counter [seg cnt in \figref{histogramModule}\panel{b}]
allows to distinguish alternating experimental scenarios, such
as when the feedback is enabled or disabled alternately in
consecutive runs of the experiment. In the correlation mode, the
value of $\tilde{Q}$ is in principle not needed but is an
additional useful piece of information. In order to make use of
the total amount of $2^{25}$ bits (4 MB) available space in the
ZBT RAM, we reduce the Q dimension to 5 bits and concatenate the
values ($\tilde{I}_2$, $\tilde{Q}$, $\tilde{I}_1$, seg) into a
21-bit address with a word size of 16 bits to store the counts
as presented in \figref{histogramModule}\panel{b}.

In the time-resolved mode, a time counter [time cnt in
\figref{histogramModule}\panel{c}] is used to add time as an
additional dimension of the histogram. The time counter starts
upon the reception of the $\fbTime$ marker and the enable flag
is held active for up to 16 clock cycles. As for the correlation
mode there is a segment counter [seg cnt in
\figref{histogramModule}\panel{c}] which allows to discern
different consecutive scenarios such as when the qubit is
prepared in the $\lvert g\rangle$ or $\lvert e\rangle$ state
alternately. This makes the time-resolved mode useful for
calibration tasks such as finding the optimal qubit readout time
by observing the separation of the distributions of $I$ and $Q$
values for the states $\lvert g\rangle$ [blue histogram in
\figref{histogramModule}\panel{d}] and $\lvert e\rangle$ [red
histogram in \figref{histogramModule}\panel{d}] over time.

\section{FPGA timing and resource analysis}
\label{app:FPGAtimingResources}

\subsection{FPGA timing analysis}\label{app:FPGAtiming}
Using the Xilinx ISE\textsuperscript{\textregistered} tool
suite~\cite{Xilinx2012}, we extracted information about the
timing of the signal processing for our present implementation
of the DSP circuit in the \hbox{\Virtex4} (xc4vsx35-10ff668) and
for future implementations on the \hbox{\Virtex6}
(xc6vlx240t-1ff1156) and \hbox{\Virtex7} (xc7vx485t-2ffg1761c)
FPGA. In \appref{FPGAresources}, we present the corresponding
FPGA resource allocations.

We define the pad--to--pad delay $\tau_\mathrm{p-p}$ as the
delay the digitized signal encounters in the path from the
signal input pads of the FPGA to the feedback trigger output
pad.
For the full implementation on the \hbox{\Virtex4} FPGA (``V--4
full'' in \tabref{FPGAtiming}), the predicted pad--to--pad delay
amounts to
\begin{align}\label{eq:padToPadV4full}
\tau_\mathrm{p-p} & \equiv \tau_\mathrm{p-f} +
\tau_\mathrm{proc} + \tau_\mathrm{f-p} \nonumber\\*
& = 1.5\ns + 30\ns + 3.8\ns = 35.3\ns,
\end{align}
where $\tauProc = 30\ns$ is the processing latency of three
pipeline stages~(see blue dashed lines in \figref{firmware}).
Moreover, the pad--to--flip--flop delay $\tau_\mathrm{p-f} =
1.5\ns$ is the maximum delay from the ADC input pads to the
D--flip--flops of the first pipelined register. Furthermore, the
flip--flop--to--pad delay $\tau_\mathrm{f-p} = 3.8\ns$ is the
maximum delay from the flip--flops of the last pipelined
register to the output pad of the feedback trigger. The
pad--to--flip--flop $\tau_\mathrm{p-f}$ and flip--flop--to--pad
$\tau_\mathrm{f-p}$ delays are expected to contribute to the
digital input and output delay $\tauADCDIO$~(see
\secref{latencies}).

A clock period analysis shows that the minimum clock period due
to the timing of the signals between two pipelined registers
amounts to $T_\mathrm{min} = 6.7\ns$ which corresponds to a
maximum clock frequency of $f_\mathrm{max} = 149\MHz$.
Increasing the clock frequency in a pipelined architecture is
however only beneficial when the sampling rate of the ADC is
also increased. Instead, removing pipeline stages in the signal
path can lead to a further decrease in processing time as long
as the minimal clock period is larger than the sampling period,
i.e. $T_\mathrm{min} \geq 1/\fs$.

As a first step towards a future optimization of the processing and
pad--to--pad delay, we separately simulated the implementation
of what we consider the core feedback functionality of the DSP
circuit which only includes the $\fs/4$ mixer, the moving
average, the offset subtraction and scaling modules and the
state discrimination module. For the implementation of the core
DSP circuit we keep only two pipelined registers, one at the ADC
input and one at the feedback outputs $\fb$ and $\fb2$.
Therefore the processing latency amounts to one clock cycle. In
order to optimize the pad-to-pad delay, we optimized first the
register-to-register delay, which determines the maximal clock
frequency. Afterwards, we optimize the pad-to-register and
register-to-pad delays. Assuming that the sampling rate is equal
to the maximal clock frequency, we obtain a pad--to--pad delay
of $4\ns+9.7\ns+7\ns = 20.7\ns$ [c.f. \eqref{padToPadV4full}]
for the \hbox{\Virtex4} implementation, $3\ns + 6.2\ns + 5.5\ns
= 14.7\ns$ [c.f. \eqref{padToPadV4full}] for the \hbox{\Virtex6}
implementation (``V--6 core'' in \tabref{FPGAtiming}), and $4\ns
+ 5.3\ns + 5\ns = 14.3\ns$ for the \hbox{\Virtex7}
implementation (``V--7 core'' in \tabref{FPGAtiming}). These
results show that a further reduction of the latency introduced
by the DSP from $35.3\ns$ to $14.3\ns$ is possible by an
optimized implementation of the core functionalities and using a
recent FPGA. We therefore consider the integration of a recent
FPGA into our experimental setup as possible future work.

\begin{table}
\caption{\label{tab:FPGAtiming} Summary of the simulated FPGA
timings: the pad--to--pad delay $\tau_\mathrm{p-p}$ from the
data input to the feedback trigger output, the processing time
$\tau_\mathrm{proc}$, the chosen clock period
$\tau_\mathrm{clk}$, the minimum clock period
$\tau_\mathrm{clk,min}$ and the maximum clock frequency
$f_\mathrm{max}$. See main text for details. }
\begin{ruledtabular}
\begin{tabular}{lrrrrr}
	&
	\parbox[c]{1.2cm}{\raggedleft$\tau_\mathrm{p-p}$ \\
		$[\mathrm{ns}]$} &
	\parbox[c]{1.2cm}{\raggedleft$\tau_\mathrm{proc}$ \\
		$[\mathrm{ns}]$} &
	\parbox[c]{1.2cm}{\raggedleft$\tau_\mathrm{clk}$ \\
		$[\mathrm{ns}]$} &
	\parbox[c]{1.2cm}{\raggedleft$\tau_\mathrm{clk,min}$ \\
		$[\mathrm{ns}]$} &
	\parbox[c]{1.2cm}{\raggedleft$f_\mathrm{max}$ \\
		$[\mathrm{MHz}]$} \\
	\hline
	       	V--4 full & 35.3 & 30 & 10 & 6.7  & 149 \\
	V--4 core & 20.7 & 10 & 9.7 & 9.7 & 103  \\
	V--6 core & 14.7 & 6.2 & 6.2 & 6.2 & 161 \\
	V--7 core & 14.3 & 5.3 & 5.3 & 5.3 & 188
\end{tabular}
\end{ruledtabular}
\end{table}

\subsection{FPGA resource analysis}\label{app:FPGAresources}

Here we report the FPGA resource allocation for the full design
implemented on the \hbox{\Virtex4} FPGA and compare it to the
resources needed to implement the core functionality consisting
of the $\fs/4$ mixer (\appref{mixerImplementation}), the moving
average (\appref{movingAverageImplementation}), the
preprocessing module
(\appref{preprocessingModuleImplementation}) and the state
discrimination module
(\appref{stateDeterminationModuleImplementation}). The analysis
of the resource allocation is done for the implementation of the
core design on the \hbox{\Virtex4}, \hbox{\Virtex6} and
\hbox{\Virtex7} FPGAs corresponding to the timing analysis
performed in \appref{FPGAtiming}.

The resource usage is summarized in \tabref{FPGAresources}. The
full design (V--4 full) uses $n_\mathrm{DFF} = 15'312$
D--flip--flops corresponding to $49\%$ of the total number of
D--flip--flops and $n_\mathrm{LUT} = 18'361$ four-input lookup
tables (LUT) which is $59\%$ of the available LUTs on the
\hbox{\Virtex4} FPGA. The majority of resources in the full
design is consumed by the flexibility in the signal processing,
such as the phase-adjustable mixer and the FIR filter with
arbitrary coefficients and the possibility to record histograms.
In addition, the full design includes hardware modules for
interfacing with PC and ZBT memory. To implement the added
flexibility in the signal processing, the full design requires
$n_\mathrm{DSP} = 184$ dedicated DSP slice resources, which
contain multipliers and adders.

\begin{table}
\caption{\label{tab:FPGAresources} FPGA resource summary
specifying the allocated number of D--flip--flops
$n_\mathrm{DFF}$, number of LUTs $n_\mathrm{LUT}$ and number of
dedicated DSP slice resources $n_\mathrm{DSP}$. Percentages are
relative (rel.) to the total amount of resources on the corresponding
FPGA. See main text for details. }
\begin{ruledtabular}
\begin{tabular}{lrr@{\hskip 5mm}rr@{\hskip 5mm}rr}
	& $n_\mathrm{DFF}$ & (rel.)
	& $n_\mathrm{LUT}$  & (rel.)
	& $n_\mathrm{DSP}$  & (rel.) \\
	\hline
	V--4 full & 15'312 & (49\%) & 18'361 & (59\%) & 184 & (95\%) \\
	V--4 core &  361     & (1\%)   &      535 & (2\%)  & 0 & \\
	V--6 core &  372     & ($<1\%$)  &       369 & ($<1\%$)  & 0 &\\
	V--7 core &  371     & ($<1\%$)  &       509 & ($<1\%$)  & 0 &
\end{tabular}
\end{ruledtabular}
\end{table}

The core design (V--4 core, V--6 core and V--7 core in
\tabref{FPGAresources}) implements only a subset of the
functionality to maintain the minimal requirements for the DSP
operations. Therefore, the number of D--flip--flops
$n_\mathrm{DFF}$ and LUTs $n_\mathrm{LUT}$ is reduced by almost
two orders of magnitude compared to the full implementation. In
addition, the implementation of the core functionality does not
require dedicated DSP slices of the FPGA ($n_\mathrm{DSP}$ in
\tabref{FPGAresources}) since no multipliers are used in the
blocks of the core design. The numbers $n_\mathrm{DFF}$ and
$n_\mathrm{LUT}$ vary depending on whether the core design is
implemented for the \hbox{\Virtex4} (\hbox{V--4} core),
\hbox{\Virtex6} (\hbox{V--6} core) or \hbox{\Virtex7}
(\hbox{V--7} core) FPGA, as displayed in \tabref{FPGAresources}.
We ascribe the variations of resource usage among the
implementations of the core design to differences in the slice
and LUT structure between the respective FPGA models. Different
slice and LUT structures result in differences of the resource
optimization in the mapping process using the Xilinx ISE
software.

\section{Experimental parameters}
\label{app:experimentalParameters}

The superconducting transmon qubit~\cite{Koch2007} has a
resonance frequency $\omega_\mathrm{q}/(2\pi) = 6.148\GHz$
corresponding to the transition between the ground and first
excited state and an anharmonicity of $\alpha=-401\MHz$. The
qubit is capacitively coupled to a $\lambda/2$ coplanar
waveguide resonator with a coupling strength
$g/(2\pi)\approx65\MHz$. We measure a fundamental mode resonance
frequency of $\omega_\mathrm{r}/(2\pi) = 7.133\GHz$ defined as
the center of the dispersively shifted resonance frequencies for
the qubit states $\lvert g\rangle$ and $\lvert e\rangle$. We
measure a linewidth of $\kappa/(2\pi) = 6.3\MHz$ of the
resonator. The qubit shows an exponential energy relaxation with
time constant $T_1\approx1.4\us$. We choose an experiment
repetition period of $10\us$, which for the given $T_1$, is
sufficient to obtain a residual out-of-equilibrium excited state
population of $0.1\%$ . The measured thermal equilibrium excited
state probability is \markred{$\Ptherm\approx7\%$}~(see
\appref{thermal}).

The envelope of the microwave pulses for qubit rotations is
Gaussian with $\sigma=7\ns$, truncated symmetrically at
$\pm2\sigma$ as seen in the pulse scheme
\figref{resultsCombined}\panel{b} and uses the DRAG
technique~\cite{Motzoi2009,Gambetta2011} to avoid errors due to
the presence of states outside the qubit subspace.

From pulsed spectroscopy we observe a dispersive shift of the
resonator frequency $\omega_{r}^{\g(\e)}$ for the qubit in the
ground $\g$ or excited state $\e$ of
\begin{equation}
2\chi \equiv \omega_\mathrm{r}^{\e}-\omega_\mathrm{r}^{\g} =
-2.2\MHz\times2\pi.
\end{equation}
We choose the frequency of the resonator drive pulses for
dispersive readout at the center between the two shifted
resonator frequencies, i.e $\omega_\mathrm{r} \equiv
(\omega_\mathrm{r}^{\e}+\omega_\mathrm{r}^{\g})/2$. The
amplitude of the readout pulse is chosen such that the expected
steady-state mean photon number is $\langle
n\rangle_\mathrm{readout}\approx10$, which we calibrated by
measuring the ac Stark shift~\cite{Schuster2005} of the qubit
frequency when a continuous coherent drive is applied to the
resonator.

\section{Reduction of thermal excited state population}
\label{app:thermal}

A possible application of active feedback initialization is to
temporarily reduce the excited state population when the qubit
is initially in thermal equilibrium with its
environment~\cite{Johnson2012,Riste2012,Riste2012b}. In order to
test the performance of our feedback loop for reducing the
thermal excited state population, we omit the $\pi/2$ pulse in
the beginning of the protocol presented in \secref{experiment},
such that the expected input state is a mixed state described by
the density matrix
\begin{equation}
	\rho_\text{therm} \equiv (1-\Ptherm)\g\gbra + \Ptherm\e\ebra,
\end{equation}
where $\Ptherm$ is the excited state population when the system
is in thermal equilibrium with its environment.

\begin{figure}
  \includegraphics{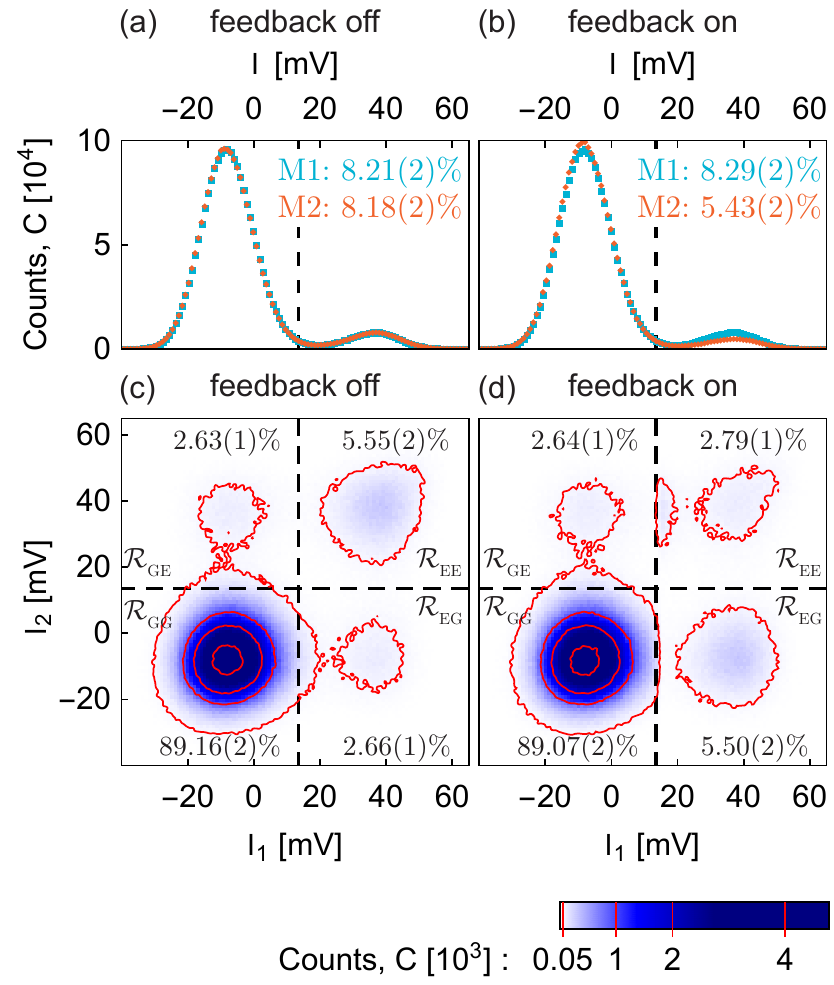}
\caption{Same type of histograms as presented in
\figref{resultsCombined} for the scenario when the initial
$\pi/2$ pulse is omitted. (a) Histograms of the in-phase signal
in the first measurement $I_1$ (blue dots) and second
measurement $I_2$ (orange dots) when feedback is disabled. The
dashed
line marks the feedback threshold. Percentages are the summed
counts of occurrences above the threshold relative to the total
count $C_\mathrm{tot} = 2'097'152$ for the signal in M1 and M2
respectively.
(b) The same type of histograms as in (a) but with feedback
enabled.
(c) Two-dimensional histogram with $128\times128$ bins as a
function of the in-phase signal in the first measurement $I_1$
versus the in-phase signal in the second measurement $I_2$ with
feedback disabled. Red lines are contour lines marking specific
counts of $\{0.05, 1, 2, 4\}\times 10^3$. In each region
$(\RGG,\RGE,\REG,\REE)$ separated by the threshold (dashed
lines) the percentage of counts relative to the total count is
indicated. (d) The same type of two-dimensional histogram as in
(c) but with feedback enabled.}
  \label{fig:resultsNoPulse}
\end{figure}

As discussed in \secref{experiment}, the qubit state is measured
by two successive readout pulses $\Mone$ and $\Mtwo$. When the
feedback actuator is disabled, the measured histogram of the
in-phase component $I_1$ of the signal during $\Mone$ (blue dots
in \figref{resultsNoPulse}\panel{a}] is almost identical to the
histogram of the in-phase component $I_2$ of the signal during
$\Mtwo$ [orange dots in \figref{resultsNoPulse}\panel{a}). From
counting the values on the right side of the threshold [dashed
line in \figref{resultsNoPulse}\panel{a}], we obtain
corresponding thermal excited state probabilities of
${\Prob[E_1]_\text{fb off} = 8.21(2)\%}$ for the first
measurement $\Mone$ and ${\Prob[E_2]_\text{fb off} = 8.18(2)\%}$
for the second measurement $\Mtwo$. This indicates that, without
conditioning on the measurement outcome, the measurement leaves
the thermal steady state $\rho_\text{therm}$ unperturbed. Note
that the overlap readout error~(see \appref{readout}) biases the
measured probabilities towards $50\%$. Taking this bias into
account, we infer a thermal excited state population of
\markred{$\Ptherm\approx7\%$} from the measured probabilities
$\Prob[E_1]$ and $\Prob[E_2]$. The inferred thermal excited
state population $\Ptherm$ corresponds to a temperature of a
bosonic environment of \markred{$T_\mathrm{env}\approx114\mK$}.
The effective temperature $T_\mathrm{env}$ is close to the
measured base temperature of the dilution refrigerator which for
the presented experiment was $90\mK$ instead of the typical
temperature of $20\mK$ due to problems with the cryogenic setup.

When feedback is enabled, the excited state probability in the
second measurement amounts to ${\Prob[E_2]_\text{fb on} =
5.43(2)\%}$, as obtained from the histogram of $I_2$ [orange
dots in \figref{resultsNoPulse}\panel{b}], is reduced compared
to the excited state probability in the first measurement
${\Prob[E_1]_\text{fb on} = 8.29(2)\%}$ obtained from the
histogram of $I_1$ [blue dots in
\figref{resultsNoPulse}\panel{b}], showing that a reduction of
the thermal excited state population is possible with our
feedback loop. The measured probability ${\Prob[E_2]_\text{fb
on}}$ is in reasonably good agreement with the simulated value
of \markred{${\Prob[E_2]_\text{fb on, sim} = 5.24\%}$} obtained
from a master equation simulation using the same model and
parameters as discussed in \secref{experiment}.

We recorded two-dimensional histograms of the values $I_1$ and
$I_2$ for the case when feedback is disabled and enabled as
shown in \figref{resultsNoPulse}\panel{c} and
\figref{resultsNoPulse}\panel{d} respectively. The measured
relative counts in the four regions \hbox{($\RGG$, $\RGE$,
$\REG$, $\REE$)} of the two-dimensional histograms show the
swapping of the probabilities ${\Prob[\REG]}$ and
${\Prob[\REE]}$ and the invariance of the probabilities
${\Prob[\RGG]}$ and ${\Prob[\RGE]}$ under the feedback action as
discussed in \secref{experiment}. The histogram of the signal in
the region $\REG$ for the "feedback off" case
[\figref{resultsNoPulse}\panel{c}] matches well with the
histogram in region $\REE$ for the "feedback on" case
[\figref{resultsNoPulse}\panel{d}]. In particular, the
corresponding probabilities ${\Prob[\REG]_\text{fb off} =
2.66(1)\%}$ and ${\Prob[\REE]_\text{fb on} = 2.79(2)\%}$ match
reasonably well, which shows that the feedback pulse is applied
when the state $\e$ is detected in the first measurement. The
experimentally observed probabilities are in reasonably good
agreement with the simulation results
\markred{${\Prob[\REG]_\text{fb off, sim} = 1.98\%}$} and
\markred{${\Prob[\REE]_\text{fb on, sim} =
1.45\%}$}~(\tabref{jointPnoPulse}) considering the sources of
systematic errors as discussed in~\secref{experiment}. We
observe that the histogram in the region $\REE$ in
\figref{resultsNoPulse}\panel{d} is double-peaked, which is a
consequence of the readout error since the tail of the
distribution associated with the $\g$ state extends into the
region $\REG$.

Furthermore, the histograms in the region $\RGE$ match for both
the "feedback off" [\figref{resultsNoPulse}\panel{c}] and the
"feedback on" case [\figref{resultsNoPulse}\panel{d}]. The
probabilities of ${\Prob[\RGE]_\text{fb off} = 2.63(1)\%}$ and
${\Prob[\RGE]_\text{fb on} = 2.64(1)\%}$ agree within the
statistical error bars, which shows that the feedback pulse is
not applied when the state $\g$ is detected in the first
measurement.

\begin{table}
\caption{\label{tab:jointPnoPulse} Experimental (exp.) and
simulated (sim.) probabilities ${\Prob[\mathcal{R}_{xy}]}$ of
the events to observe the qubit in state~$x$ in the first
measurement and in state~$y$ in the second measurement when the
feedback is either disabled or enabled. The simulated values are
obtained from a master equation simulation. See text for
details.}
    \begin{ruledtabular}
        \begin{tabular}{l|rr|rr}
& \multicolumn{2}{c|}{feedback off} &
\multicolumn{2}{c}{feedback on}\\
             & exp. & sim. & exp. & sim. \\
            \hline
$\Prob[\RGG]$ & 89.16(2)\% & \markred{88.01\%} & 89.07(2)\% &
\markred{88.01\%} \\
$\Prob[\RGE]$ & 2.63(1)\% & \markred{3.79\%} & 2.64(1)\% &
\markred{3.79\%} \\
$\Prob[\REG]$ & 2.66(1)\% & \markred{1.98\%} & 5.50(2)\% &
\markred{6.75\%} \\
$\Prob[\REE]$ & 5.55(2)\% & \markred{6.22\%} & 2.79(1)\% &
\markred{1.45\%} \\
        \end{tabular}
    \end{ruledtabular}
\end{table}

The data presented here serves as a further experimental
benchmark of our implementation of the feedback scheme and
illustrates the use of two-dimensional histograms to get insight
into processes that lead to the observed excited state
probabilities.

\section{Readout fidelity}\label{app:readout}

The readout is calibrated in a separate calibration step where
either no pulse or a $\pi$ pulse is applied to the qubit prior
to the measurement. A threshold check, as described in
\secref{readoutScheme}, leads either to the result $G$
corresponding the qubit state $\g$ or $E$ corresponding to $\e$.
The single-shot readout fidelity is defined as
\begin{equation}
F_\mathrm{r}= 1- \Prob[E|\text{``no
pulse"}]-\Prob[G|\text{``$\pi$ pulse"}],
\end{equation}
where ${\Prob[E|\text{``no pulse"}]}$ represents the conditional
probability of obtaining the result $E$ when no pulse has been
applied whereas ${\Prob[G|\text{``$\pi$ pulse"}]}$ represents
the conditional probability of obtaining the result $G$ when a
$\pi$ pulse has been issued.
For a fixed moving average window length of $l = 4$ digital
samples ($40\ns$), the single-shot fidelity reaches a maximal
value of $F_\mathrm{r} = 77\%$ at time
\markred{$\tauRO\approx105\ns$} relative to the onset of the
readout pulse. We expect the contributions to the readout
infidelity to be
\begin{equation}
1-F_\mathrm{r} \approx 2\Ptherm + P_{\text{decay}} +
P_{\text{overlap}},
\end{equation}
where \markred{$\Ptherm\approx7\%$} is the initial excited
state population in thermal equilibrium~(see \appref{thermal}),
${P_{\text{decay}} \approx 1-\exp(-\tauRO/T1) \approx 6\%}$ is
the error due to the decay of the $\e$ state and
$P_{\text{overlap}}\approx3\%$ is the probability of
misidentification of the qubit state due to overlap of the
probability density functions for the signals corresponding to
the $\g$ and $\e$ state. We extracted the contributions to the
readout infidelity from fits to recorded histograms using
methods similar to the ones described in~\citeref{Walter2017}.


\bibliography{Q:/USERS/Yves/_literature/RefDB/QudevRefDB}

\end{document}